\documentclass[11pt,a4paper]{article}

\usepackage{jheppub}
\usepackage[utf8]{inputenc}
\usepackage[english]{babel}
\usepackage{mathrsfs}
\usepackage{slashed}
\usepackage{verbatim}
\usepackage{multirow}
\usepackage{empheq}
\usepackage{arydshln}

\newcommand{\wti}[1]{\ensuremath{\widetilde{#1}}}
\newcommand{\sca}[1]{\ensuremath{\left(#1\right)}}
\renewcommand{\b}[1]{\boldsymbol{#1}}
\newcommand*{\vv}[1]{\vec{\mkern0mu#1}}

\begin{document}

\title{Renormalization group invariants in supersymmetric theories: one- and two-loop results}

\author[a,b]{Wim Beenakker,}
\author[c,d]{Tom van Daal,}
\author[a]{Ronald Kleiss,}
\author[a]{Rob Verheyen}

\affiliation[a]{Institute for Mathematics, Astrophysics and Particle Physics, Faculty of Science, Mailbox 79, Radboud University Nijmegen, P.O. Box 9010, 6500 GL Nijmegen, The Netherlands}
\affiliation[b]{Institute of Physics, University of Amsterdam, Science Park 904, 1018 XE Amsterdam, The Netherlands}
\affiliation[c]{Department of Physics and Astronomy, VU University Amsterdam, De Boelelaan 1081, 1081 HV Amsterdam, The Netherlands}
\affiliation[d]{Nikhef, Science Park 105, 1098 XG Amsterdam, The Netherlands}

\emailAdd{W.Beenakker@science.ru.nl}
\emailAdd{tvdaal@nikhef.nl}
\emailAdd{R.Kleiss@science.ru.nl}
\emailAdd{robverheyen@gmail.com}

\abstract{We stress the potential usefulness of renormalization group invariants. Especially particular combinations thereof could for instance be used as probes into patterns of supersymmetry breaking in the MSSM at inaccessibly high energies. We search for these renormalization group invariants in two systematic ways: on the one hand by making use of symmetry arguments and on the other by means of a completely automated exhaustive search through a large class of candidate invariants. At the one-loop level, we find all known invariants for the MSSM and in fact several more, and extend our results to the more constrained pMSSM and dMSSM, leading to even more invariants. Extending our search to the two-loop level we find that the number of invariants is considerably reduced.}


\maketitle

\section{Introduction}
It has long been known that the Standard Model cannot be the final theory of particle physics. Issues such as the hierarchy problem or the absence of a satisfactory description of dark matter or gravity lead theoretical physicists to develop more fundamental theories. Supersymmetry is one of those theories, and the \emph{Minimal Supersymmetric Standard Model} (MSSM) will be one of the main focuses for the second run of the Large Hadron Collider. If signs of supersymmetry are indeed found, theorists will face the issue of figuring out the exact underlying theoretical description. In case of the MSSM, this problem includes finding the exact mechanism of supersymmetry breaking. If supersymmetry is a symmetry of nature, every particle must have the same mass as its superpartner. Since this is clearly not the case, supersymmetry must be broken. The MSSM accounts for several breaking mechanisms by incorporating all possible soft supersymmetry breaking terms in its Lagrangian. If any of the several possible breaking mechanisms is realized in nature, this will be signified by a characteristic unification of some soft supersymmetry breaking parameters at a high energy scale.

These unifications can be studied through renormalization group (RG) techniques. Typical strategies include evolving the values of measured parameters upward to the unification scale (bottom-up), or choosing values of the parameters at the unification scale that are evolved downward (top-down). We will discuss a third method that makes use of RG invariant combinations of RG equations, called \emph{RG invariants}. Using RG invariants to probe high-scale physics, has several distinct advantages over the bottom-up and top-down methods \cite{Demir:2004aq,Carena:2010gr,Hetzel:2012bk}. Finding these invariants can be very difficult though. In this paper we will discuss two different methods to find RG invariants. One method relates invariants to symmetries of the underlying theory and the other is based on computer algebraic techniques. Both methods will be applied to the MSSM, the diagonal MSSM (dMSSM), and the phenomenological MSSM (pMSSM) to find invariants up to two-loop order.

The paper is organized as follows. In section \ref{2} we describe the method of RG invariants and its merits. In section \ref{3} we investigate whether some of these RG invariants can be related to symmetries of the underlying theory. For this underlying theory we will consider the MSSM, dMSSM, and pMSSM. In section \ref{4} a generic computer algebraic method for finding RG invariants is 
introduced. This method is used to find previously unknown invariants at one- and two-loop level within the aforementioned supersymmetric theories.

\section{Probing high-scale physics with RG invariants} \label{2}
Let us consider a renormalized theory with a running parameter $p(\mu)$. We define the corresponding $\beta$-function as follows:
\begin{equation} \label{2.1}
\beta(p) \equiv 16 \pi^2 \frac{dp}{dt} ,
\end{equation}
where $t \equiv \log_{10}(\mu/\mu_{0})$. The energy scale $\mu$ is normalized by an arbitrary reference scale $\mu_{0}$ to make the logarithm dimensionless. For quantum field theories, the one-loop $\beta$-functions are polynomials of the parameters of the theory with rational coefficients. Higher order contributions contain additional factors of $1/(16\pi^2)$.

How could RG invariants be used to probe high-scale physics? We will explain this through a toy system of one-loop $\beta$-functions for the parameters $v,w,x,y,z$ that closely resembles some of the one-loop MSSM $\beta$-functions; we define
\begin{subequations}
\begin{align}
\beta(v) &= v^{3} \label{2a} , \\
\beta(w) &= v^{2}\left(5w+6x-4y\right) , \\
\beta(x) &= v^{2}\left(-w-2x+4z\right) , \\
\beta(y) &= v^{2}\left(x+y-5z\right) , \\
\beta(z) &= v^{2}\left(w-2y+6z\right) \label{2e} .
\end{align}
\end{subequations}
This system of $\beta$-functions was built such that the parameter $v$ resembles a gauge coupling parameter, while $w,x,y,z$ resemble MSSM scalar masses. An RG invariant is an algebraic combination of parameters $I$, such that
\begin{equation} \label{2.3}
\frac{d}{dt}I=0 .
\end{equation}

As it turns out, two independent RG invariants can be constructed from the $\beta$-functions given by eqs. \eqref{2a} - \eqref{2e}; we define
\begin{equation} \label{2.4}
I_1 \equiv w+3x-2z , \quad I_2 \equiv x+2y+z .
\end{equation}
Supersymmetry breaking mechanisms typically predict the unification of scalar masses at some large, experimentally inaccessible energy scale. We can use the invariants in \eqref{2.4} to check whether the unification of $w,x,y,z$ is realized in nature. Suppose that these scalar masses unify to the value $s$ at some scale, then we would have
\begin{equation} 
I_1=2s , \quad I_2=4s ,
\end{equation}
from which it follows that 
\begin{equation} \label{2.5}
2I_1-I_2=0 .
\end{equation}
Such a relation between RG invariants is called a \emph{sum rule}. Since $I_1$ and $I_2$ are invariant under RG flow, sum rules such as eq. \eqref{2.5} remain true at all energy scales, \emph{if} unification occurs. In particular, sum rules can be checked at the collider scale, potentially falsifying the assumed unification. 

The approach using RG invariants avoids a number of issues that the top-down and bottom-up methods suffer from. The top-down method requires knowledge of both the unification scale and value in order to evolve the parameters down to experimentally accessible scales. Since theory does not predict these values with much accuracy, scans are typically performed over a range of scales and values. These scans can be computationally very time-consuming (depending on the number of unifying parameters), and are often not feasible. The use of RG invariants requires no knowledge of the unification scale or value whatsoever, so that this problem is avoided.

\begin{figure}[!th]
\centering
\includegraphics[width=\textwidth]{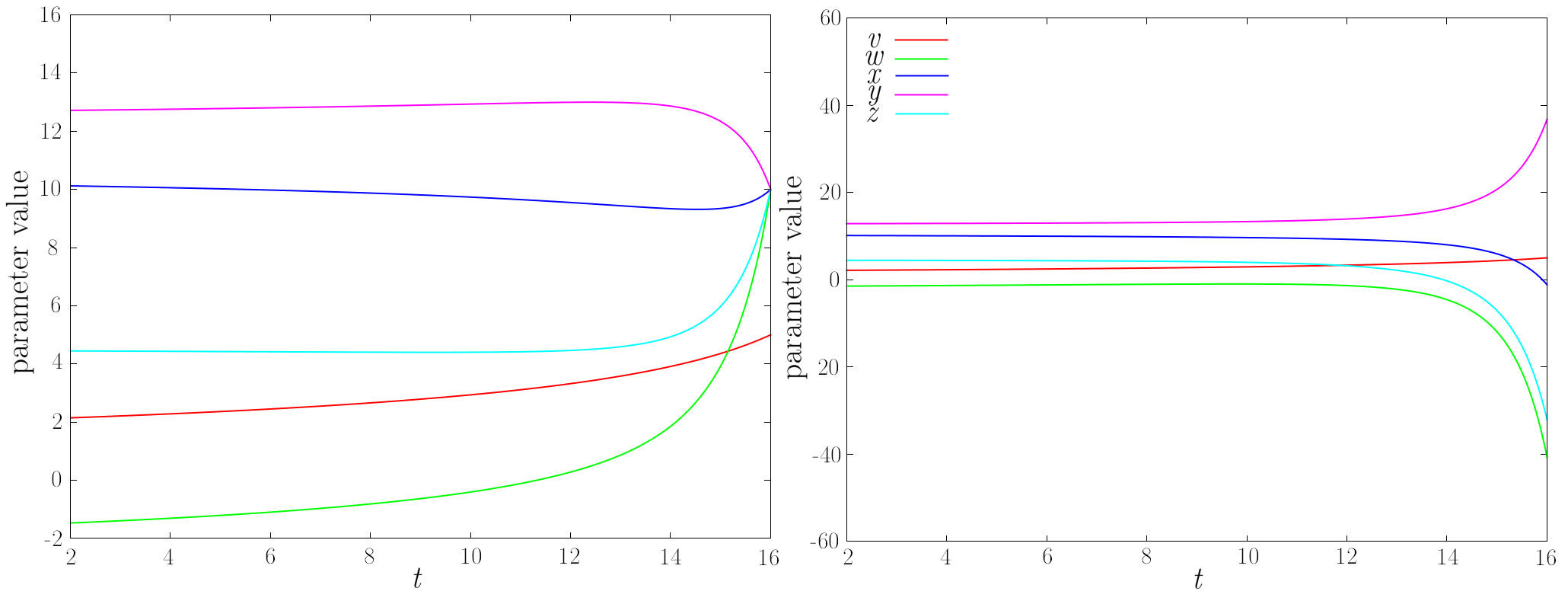}
\caption{Left: The evolution of the parameters $v,w,x,y,z$ according to their one-loop $\beta$-functions \eqref{2a} - \eqref{2e}. The parameters $w,x,y,z$ unify at $t=16$ to the value $10$. Right: The initial
value of $y$ at $t=2$ is raised by $1\%$.}
\label{Fig1}
\end{figure}

The bottom-up method does not require knowledge of unifying scales or values either, but suffers from a different problem resulting from the numerical evolution of experimental input values. When parameters are evolved up to higher scales, any experimental errors are typically greatly enhanced. Figure \ref{Fig1} shows the running effects of a slight change of one of the low-scale values of the parameters $v,w,x,y,z$ when $w,x,y,z$ unify at some high scale. After the slight change, the evolution of the parameters no longer shows any sign of unification whatsoever. The use of RG invariants circumvents such numerical problems. \\

Finally,  it is usually not necessary to know all parameters of the theory when using RG-invariants as probes of high-scale physics. The $\beta$-functions for most theories are heavily coupled differential equations, and hence evolving the relevant parameters of a theory usually requires the evolution of all other parameters of the theory as well. In our example the sum rule of eq. \eqref{2.5} does not involve the parameter $v$, which does not participate in the unification anyway. Both the top-down and bottom-up methods also require a value for $v$ to numerically evolve the other parameters. 

Constructing RG invariants is generally a very non-trivial matter. To find invariants, we will not use the ``standard'' algebraical techniques that are used in for example \cite{Hetzel:2012bk}. Instead, we will consider two different, more efficient methods: one relies on symmetries of the underlying theory, and the other is based on computer algebraic techniques. Both methods will be applied to find invariants for the MSSM, the dMSSM, and the pMSSM.

\section{RG invariants from symmetries} \label{3}
First, we will look at the construction of RG invariants from a perspective that involves symmetries of the underlying theory. The existence of such a relation between symmetries and invariants was suggested by \cite{Carena:2010gr} in the context of the pMSSM. As yet, no compelling arguments or proof of this has been presented though. First we will consider the MSSM, then we will study a ``flavor-diagonal'' version of the MSSM (the dMSSM), and, finally, the pMSSM will be discussed. All nomenclature regarding MSSM, dMSSM, and pMSSM fields and parameters is defined in appendices \ref{A}, \ref{B}, and \ref{C} respectively.

\subsection{The MSSM}
In this subsection we attempt to construct RG invariants in the MSSM. So far, RG invariants have only been constructed for heavily constrained supersymmetric models, such as the pMSSM \cite{Carena:2010gr,Hetzel:2012bk}. We will attempt to construct invariants for the full 168-parameter MSSM.

\subsubsection{Known invariants}
We expect that some invariants that have been found in the pMSSM also exist in the MSSM, since the simplifications of the pMSSM with respect to the MSSM mainly apply to the family sector, and, for example, not at all to the gauge sector. The following invariants that have been found in the pMSSM are also invariants in the MSSM, which can be checked easily with the $\beta$-functions in appendix \ref{A.4}. In the combined gauge and gaugino sectors, we can construct three invariants:\footnote{Strictly speaking there are six invariants: these three, plus their complex conjugates.}
\begin{equation}
I_1 \equiv \frac{M_1}{g'^2} , \quad I_2 \equiv \frac{M_2}{g^2} , \quad I_3 \equiv \frac{M_3}{g_\text{s}^2} ,
\end{equation}
and in the pure gauge sector we have
\begin{equation}
I_4 \equiv \frac{1}{g'^2} - \frac{11}{g^2} , \quad I_5 \equiv \frac{3}{g'^2} + \frac{11}{g_\text{s}^2} .
\end{equation}
Finally, the invariant that involves the quantity $S$ turns out to be an invariant of the MSSM as well:
\begin{equation}
I_6 \equiv \frac{S}{g'^2} .
\label{Sg'}
\end{equation}
Thus, six invariants that we already know from the pMSSM carry over trivially to the MSSM. 

\subsubsection{New invariants}
As we know from \cite{Carena:2010gr,Hetzel:2012bk}, all remaining invariants in the pMSSM (there exist eight more besides $I_1, \ldots, I_6$) involve scalar masses only, or combinations of scalar and gaugino masses.\footnote{Actually, in the pMSSM two more invariants exist that involve the Higgs mixing parameters $\mu$ and $b$, but as argued in \cite{Hetzel:2012bk} these are useless to probe high-scale physics models. For completeness, though, they are listed in appendix \ref{D}.} For this reason, we will now focus on these sectors in the MSSM to see if more invariant quantities can be constructed.

Let us now try to construct invariants from the one-loop $\beta$-functions of the soft scalar masses and gaugino masses that are listed in appendix \ref{A.4}. The first thing we note when we look at the $\beta$-functions for the soft scalar masses, is that no invariants can possibly be constructed for the off-diagonal terms of the sfermion mass matrices. The reason for this is that the order in which the family space matrices (i.e. the Yukawa and trilinear coupling matrices) appear is different for all five $\beta$-functions of the sfermion mass matrices. In other words, there are simply too many different structures present. In order to avoid this, we need to work with terms that are insensitive to this order of matrices. Hence, a logical step would be to consider the $\beta$-functions for the \emph{traces} of the sfermion mass matrices. These $\beta$-functions are given by
\begin{subequations}
\begin{align}
\beta \left[ \text{Tr}(\b{m_{\wti{Q}}^2}) \right] =& - 24 Y_{\wti{Q}_L}^2 g'^2 |M_1|^2 - 18 g^2 |M_2|^2 - 32 g_\text{s}^2 |M_3|^2 + 6 Y_{\wti{Q}_L} g'^2 S \label{TrQ} \nonumber \\
& + 2 \,\text{Tr} \left( m_{H_u}^2 \b{y_u^\dag y_u} + \b{m_{\wti{Q}}^2 y_u^\dag y_u} + \b{y_u^\dag m_{\wti{u}}^2 y_u} + \b{a_u^\dag a_u} \right) \nonumber \\
& + 2 \,\text{Tr} \left( m_{H_d}^2 \b{y_d^\dag y_d} + \b{m_{\wti{Q}}^2 y_d^\dag y_d} + \b{y_d^\dag m_{\wti{d}}^2 y_d} + \b{a_d^\dag a_d} \right) , \\[1mm]
\beta \left[ \text{Tr}(\b{m_{\wti{L}}^2}) \right] =& - 24 Y_{\wti{L}_L}^2 g'^2 |M_1|^2 - 18 g^2 |M_2|^2 + 6 Y_{\wti{L}_L} g'^2 S \nonumber \\
& + 2 \,\text{Tr} \left( m_{H_d}^2 \b{y_e^\dag y_e} + \b{m_{\wti{L}}^2 y_e^\dag y_e} + \b{y_e^\dag m_{\wti{e}}^2 y_e} + \b{a_e^\dag a_e} \right) , \\[1mm]
\beta \left[ \text{Tr}(\b{m_{\wti{u}}^2}) \right] =& - 24 Y_{\wti{u}_R^*}^2 g'^2 |M_1|^2 - 32 g_\text{s}^2 |M_3|^2 + 6 Y_{\wti{u}_R^*} g'^2 S \nonumber \\
& + 4 \,\text{Tr} \left( m_{H_u}^2 \b{y_u^\dag y_u} + \b{m_{\wti{Q}}^2 y_u^\dag y_u} + \b{y_u^\dag m_{\wti{u}}^2 y_u} + \b{a_u^\dag a_u} \right) , \\[1mm]
\beta \left[ \text{Tr}(\b{m_{\wti{d}}^2}) \right] =& - 24 Y_{\wti{d}_R^*}^2 g'^2 |M_1|^2 - 32 g_\text{s}^2 |M_3|^2 + 6 Y_{\wti{d}_R^*} g'^2 S \nonumber \\
& + 4 \,\text{Tr} \left( m_{H_d}^2 \b{y_d^\dag y_d} + \b{m_{\wti{Q}}^2 y_d^\dag y_d} + \b{y_d^\dag m_{\wti{d}}^2 y_d} + \b{a_d^\dag a_d} \right) , \\[1mm]
\beta \left[ \text{Tr}(\b{m_{\wti{e}}^2}) \right] =& - 24 Y_{\wti{e}_R^*}^2 g'^2 |M_1|^2 + 6 Y_{\wti{e}_R^*} g'^2 S \nonumber \\
& + 4 \,\text{Tr} \left( m_{H_d}^2 \b{y_e^\dag y_e} + \b{m_{\wti{L}}^2 y_e^\dag y_e} + \b{y_e^\dag m_{\wti{e}}^2 y_e} + \b{a_e^\dag a_e} \right) , \label{Tre} \\[1mm]
\beta(m_{H_u}^2) =& - 8 Y_{H_u}^2 g'^2 |M_1|^2 - 6 g^2 |M_2|^2 + 2 Y_{H_u} g'^2 S \nonumber \\
& + 6 \,\text{Tr} \left( m_{H_u}^2 \b{y_u^\dag y_u} + \b{m_{\wti{Q}}^2 y_u^\dag y_u} + \b{y_u^\dag m_{\wti{u}}^2 y_u} + \b{a_u^\dag a_u} \right) , \\[1mm]
\beta(m_{H_d}^2) =& - 8 Y_{H_d}^2 g'^2 |M_1|^2 - 6 g^2 |M_2|^2 + 2 Y_{H_d} g'^2 S \nonumber \\
& + 6 \,\text{Tr} \left( m_{H_d}^2 \b{y_d^\dag y_d} + \b{m_{\wti{Q}}^2 y_d^\dag y_d} + \b{y_d^\dag m_{\wti{d}}^2 y_d} + \b{a_d^\dag a_d} \right) \nonumber \\
& + 2 \,\text{Tr} \left( m_{H_d}^2 \b{y_e^\dag y_e} + \b{m_{\wti{L}}^2 y_e^\dag y_e} + \b{y_e^\dag m_{\wti{e}}^2 y_e} + \b{a_e^\dag a_e} \right) \label{mhd2} ,
\end{align}
\end{subequations}
where, for convenience, we have also added the $\beta$-functions for $m_{H_u}^2$ and $m_{H_d}^2$ (which already involved traces).

In the MSSM, \emph{all} interactions in family space are described by the Yukawa terms in the superpotential and the soft supersymmetry breaking trilinear terms (cf. appendix \ref{A.2}). In case that the order of the family space matrices does not matter by taking a trace, then these interactions give rise to three different trace structures in \eqref{TrQ} - \eqref{mhd2}, labeled by $u,d,$ and $e$. These trace structures each belong to one of the three Yukawa and trilinear interaction terms in the superpotential and the soft breaking Lagrangian respectively, and each of those terms involves a unique set of three scalar fields. For example, the terms in the MSSM Lagrangian that give rise to the trace structure that features the label $u$ are
\begin{equation}
\Delta W_\text{MSSM} = \wti{u}_R^\dag \b{y_u} (\wti{Q}_L)^\alpha (H_u)_\alpha , \quad \Delta \mathscr{L}_\text{tril.} = - \wti{u}_R^\dag \b{a_u} (\wti{Q}_L)^\alpha (H_u)_\alpha  + \text{h.c.}
\label{DeltaW}
\end{equation}

Note that the coefficients of the three trace structures are different for the various scalar masses. This is because in the $\beta$-functions no traces have been carried out over the gauge degrees of freedom of the corresponding scalar fields. How is this to be understood? Let us consider the trilinear interaction between the fields $\wti{u}_R$, $\wti{Q}_L$, and $H_u$, as well as the one-loop corrections to the scalar propagators that this interaction gives rise to. These three fields can all occur as external fields or inside loops. The gauge degrees of freedom that are ``closed'' inside the loops (i.e. those gauge degrees of freedom that the external fields do not possess) are summed over. Say we take $\wti{u}_R$ to be the external field, then there is an $\text{SU}(2)$ doublet degree of freedom inside the loop that has to be ``traced over'', giving a factor of 2. If we had taken $H_u$ to be the external field, then a trace over $\text{SU}(3)$ degrees of freedom inside the loop would have resulted, giving a factor of 3, etc. Hence, if we multiply the $\beta$-functions \eqref{TrQ} - \eqref{mhd2} by the factors that result from summing over the gauge degrees of freedom of the external scalar fields, then all three trace structures each get exactly the same coefficients. 

How many RG invariants do we expect to find in the soft scalar and gaugino sectors? We have ten equations (seven $\beta$-functions for the scalar masses and three for the gaugino masses) that contain seven different structures (three trace structures, three gaugino masses and $S$). Having ten equations to eliminate seven different terms should give $10-7=3$ independent RG invariants. 

To construct invariant quantities from these $\beta$-functions, we can first try to get rid of the three different trace structures. To cancel these structures, we could consider a linear sum of the $\beta$-functions \eqref{TrQ} - \eqref{mhd2}, appropriately multiplied by numbers of gauge degrees of freedom, such that the coefficients of this sum add up to zero for each trace structure. Now we can use the fact that each trace structure corresponds to a unique combination of three scalar fields in the Lagrangian. For the cancellation to take place, we should assign quantum numbers to these groups of three fields that each add up to zero, which is equivalent to saying that the quantum number should be conserved by all family space interactions. This means that to cancel the Yukawa and trilinear contributions to the $\beta$-functions, we should consider $\text{U}(1)$ symmetries of the MSSM family sector. More specifically, for any quantum number $Q$ that pertains to a symmetry $\text{U}(1)_Q$ of the MSSM family sector, the $\beta$-function of the quantity
\begin{equation}
\text{Tr} \bigg( \sum_\phi Q_\phi \b{m_\phi^2} \bigg) ,
\label{sum}
\end{equation}
with the sum running over all scalar fields $\phi$, does no longer contain the three trace structures. Note that sums like the one in \eqref{sum} over fields that occur inside family space traces are implicitly understood to be over all gauge degrees of freedom of the gauge multiplets.\footnote{For the field $\wti{Q}_L$, for example, such a sum would give a factor of 2 for $\text{SU}(2)$ and a factor of 3 for $\text{SU}(3)$ degrees of freedom, yielding a total multiplication by 6.} 

What could the quantum number $Q$ be? As is discussed in appendix \ref{A.3}, the relevant $\text{U}(1)$ symmetries correspond to the quantum numbers weak hypercharge ($Y$), baryon number ($B$), lepton number ($L$), and $X$. Table \ref{quantum numbers} provides the quantum numbers $Y,B,L,X$ for all scalar fields in the MSSM. The presence of exactly \emph{four} independent $\text{U}(1)$ symmetries in the family sector of the MSSM can be explained as follows: from table \ref{quantum numbers} we infer that this sector of the theory is constituted by seven scalar multiplets. We can regard each set of quantum numbers pertaining to a given symmetry as being a vector in a seven-dimensional vector space. This vector is subject to three independent symmetry constraints coming from the interaction terms in the superpotential and the soft trilinear terms (the terms in eq. \eqref{DeltaW} give one of these constraints). This means that we can construct $7-3=4$ linearly independent vectors in this space, i.e. four sets of quantum numbers each corresponding to a different $\text{U}(1)$ symmetry. \\
\begin{table}[!th]
\begin{center}
{\renewcommand{\arraystretch}{1.2}
\begin{tabular}{|c|c|c|c|c|}
\hline
\textbf{Spin 0} & $\b{Y}$ & $\b{B}$ & $\b{L}$ & $\b{X}$ \\
\hline \hline
$\wti{Q}_L$ & $\frac{1}{6}$ & $\frac{1}{3}$ & $0$ & $1$ \\ 
$\wti{L}_L$ & $-\frac{1}{2}$ & $0$ & $1$ & $1$ \\ 
$\wti{u}_R^*$ & $-\frac{2}{3}$ & $-\frac{1}{3}$ & $0$ & $1$ \\
$\wti{d}_R^*$ & $\frac{1}{3}$ & $-\frac{1}{3}$ & $0$ & $1$ \\
$\wti{e}_R^*$ & $1$ & $0$ & $-1$ & $1$ \\
$H_u$ & $\frac{1}{2}$ & $0$ & $0$ & $-2$ \\
$H_d$ & $-\frac{1}{2}$ & $0$ & $0$ & $-2$ \\
\hline
\end{tabular}}
\caption{The quantum numbers $Y,B,L,X$ for all MSSM scalar fields.}
\label{quantum numbers}
\end{center}
\end{table}

We now know how to get rid of the three family space trace structures in eqs. \eqref{TrQ} - \eqref{mhd2}, but what about the other terms present? Can quantum numbers also be used to eliminate the remaining structures (i.e. $S$ and the three absolute squared gaugino masses $|M_1|^2,|M_2|^2,|M_3|^2$)? For any $Q$, we have
\begin{align}
\beta \bigg[ \text{Tr} \bigg( \sum_\phi Q_\phi \b{m_\phi^2} \bigg) \bigg] =& \;2 \bigg( \sum_\phi Y_\phi Q_\phi \bigg) g'^2 S - 8 \bigg( \sum_\phi Y_\phi^2 Q_\phi \bigg) g'^2 |M_1|^2 \nonumber \\
& - 6 \bigg( \sum_d Q_d \bigg) g^2 |M_2|^2 - \frac{32}{3} \bigg( \sum_t Q_t \bigg) g_\text{s}^2 |M_3|^2 ,
\label{beta}
\end{align}
where $d$ denotes the scalar weak isospin doublets and $t$ the scalar color triplets. Note that sums without any family space traces involved are implicitly understood to be over all families as well (besides the gauge degrees of freedom).\footnote{This simply amounts to an additional factor of $3$ for all sfermions.} From eq. \eqref{beta} it directly follows that to eliminate $S,|M_1|^2,|M_2|^2,|M_3|^2$ respectively, we must have:
\begin{equation}
\sum_\phi Y_\phi Q_\phi = 0 , \quad \sum_\phi Y_\phi^2 Q_\phi = 0 , \quad \sum_d Q_d = 0 , \quad \sum_t Q_t = 0 .
\label{reqs}
\end{equation}
The latter three sums over charges are reminiscent of mixed anomaly cancellations of the charge $Q$ with the separate gauge groups that we know from for example the Standard Model. 

Which quantum numbers actually satisfy the requirements in \eqref{reqs}? This is summarized in table \ref{charges}. From this table it follows that to eliminate $S$, suitable quantum numbers would be $3B+L$, $11(B-L)-8Y$, and $X$. For the cancellation of $|M_1|^2$, we could use $Y$, $B-L$, or $16B + 3X$, while for $ |M_2|^2$ the quantum numbers $Y$, $B-L$, and $8B - 3X$ are suitable. For $|M_3|^2$ to cancel, we could pick $Y$, $B$, or $L$. Naturally, linear combinations of these quantum numbers also work. \\
\begin{table}[!th]
\begin{center}
{\renewcommand{\arraystretch}{1.2}
\begin{tabular}{|c|c|c|c|c|c|}
\hline
\textbf{Quantity} & \textbf{Sum} & $\b{Y}$ & $\b{B}$ & $\b{L}$ & $\b{X}$ \\
\hline \hline
$S$ & $\sum_\phi Y_\phi Q_\phi$ & $11$ & $2$ & $-6$ & $0$ \\ 
$|M_1|^2$ & $\sum_\phi Y_\phi^2 Q_\phi$ & $0$ & $-\frac{3}{2}$ & $-\frac{3}{2}$ & $8$ \\ 
$|M_2|^2$ & $\sum_d Q_d$ & $0$ & $6$ & $6$ & $16$ \\
$|M_3|^2$ & $\sum_t Q_t$ & $0$ & $0$ & $0$ & $36$ \\
\hline
\end{tabular}}
\caption{The evaluation of the sums that are related to the elimination requirements  \eqref{reqs} for the quantities $S$, $|M_{1,2,3}|^{2}$, given for the quantum numbers $Y,B,L,X$. The outcome $0$ for a certain quantum number indicates that this quantum number is suitable for eliminating the corresponding quantity from the $\beta$-functions \eqref{TrQ} - \eqref{mhd2}.} 
\label{charges}
\end{center}
\end{table}

From eq. \eqref{beta}, as well as from the $\beta$-functions for the gaugino masses \eqref{M's} and $S$ \eqref{betaS}, it follows that \emph{any} one-loop RG invariant $I$ in the MSSM that involves scalar masses and gaugino masses, is of the following form:
\begin{align}
I_Q =& \;\text{Tr} \bigg( \sum_\phi Q_\phi \b{m_\phi^2} \bigg) - \frac{1}{11} \sum_\phi Y_\phi Q_\phi S + \frac{2}{11} \sum_\phi Y_\phi^2 Q_\phi |M_1|^2 + \frac{3}{2} \sum_d Q_d |M_2|^2 \nonumber \\
& - \frac{8}{9} \sum_t Q_t |M_3|^2 ,
\label{I}
\end{align}
where $Q$ is \emph{any} quantum number that is preserved by all MSSM family space interactions (i.e. $Q$ must be a linear combination of $Y,B,L,X$). 

Now we are ready to construct RG invariants using table \ref{charges} and eq. \eqref{I}. From table \ref{charges} we infer that the quantum number $11(B-L)-8Y$ cancels $S$ \emph{and} all gaugino masses, as this linear combination of $Y,B,L$ vanishes for each row in the table. This leads us to define the following RG invariant:
\begin{align}
I_7 \equiv& \;\text{Tr} \bigg( \sum_\phi \left( 11B_\phi -11 L_\phi - 8 Y_\phi \right) \b{m_\phi^2} \bigg) \nonumber \\
=& \;\text{Tr} \left( 14 \b{m_{\wti{Q}}^2} - 14 \b{m_{\wti{L}}^2} + 5 \b{m_{\wti{u}}^2} - 19 \b{m_{\wti{d}}^2} + 3 \b{m_{\wti{e}}^2} \right) - 8 m_{H_u}^2 + 8 m_{H_d}^2 .
\end{align}
If we wish to construct an invariant where both $S$ and $|M_3|^2$ are eliminated by a symmetry, then we could use the quantum number $3B + L$. We define
\begin{align}
I_8 \equiv& \;\text{Tr} \bigg( \sum_\phi (3 B_\phi + L_\phi) \b{m_\phi^2} \bigg) + \frac{2}{11} \sum_\phi Y_\phi^2 (3 B_\phi + L_\phi) |M_1|^2 + \frac{3}{2} \sum_d (3 B_d + L_d) |M_2|^2 \nonumber \\
=& \;\text{Tr} \left( 6 \b{m_{\wti{Q}}^2} + 2 \b{m_{\wti{L}}^2} - 3 \b{m_{\wti{u}}^2} - 3 \b{m_{\wti{d}}^2} - \b{m_{\wti{e}}^2} \right) - \frac{12}{11} |M_1|^2 + 36 |M_2|^2 .
\end{align}
For the third and last \emph{independent} invariant in this sector, let us pick the quantum number $X$ (which only cancels $S$) and define\footnote{Even though we have four symmetries at hand, only three independent RG invariants can be constructed. This is because taking $Q \propto Y$ gives $I_Q = 0$. In fact, the quantum number $Y$ has already been used for the invariant that involves the quantity $S$ (cf. eq. \eqref{Sg'}).}
\begin{align}
I_9 \equiv& \;\text{Tr} \bigg( \sum_\phi X_\phi \b{m_\phi^2} \bigg) + \frac{2}{11} \sum_\phi Y_\phi^2 X_\phi |M_1|^2 + \frac{3}{2} \sum_d X_d |M_2|^2 - \frac{8}{9} \sum_t X_t |M_3|^2 \nonumber \\
=& \;\text{Tr} \left( 6 \b{m_{\wti{Q}}^2} + 2 \b{m_{\wti{L}}^2} + 3 \b{m_{\wti{u}}^2} + 3 \b{m_{\wti{d}}^2} + \b{m_{\wti{e}}^2} \right) - 4 m_{H_u}^2 - 4 m_{H_d}^2 + \frac{16}{11} |M_1|^2 \nonumber \\[2mm]
& + 24 |M_2|^2 - 32 |M_3|^2 .
\end{align}

\subsection{The dMSSM} \label{dMSSM}
Let us consider a constrained version of the MSSM, the so-called dMSSM, where all matrices in family space (i.e. the sfermion mass matrices and the Yukawa and trilinear coupling matrices) are taken diagonal (cf. appendix \ref{B} for a more extensive discussion of the dMSSM), and let us again focus on the soft scalar and gaugino sectors to find RG invariants. In this particular model, we have twenty equations (fifteen $\beta$-functions for the sfermion masses, two for the Higgs masses, and three for the gaugino masses) containing thirteen different structures (nine structures coming from the diagonal components of the Yukawa and trilinear coupling matrices, three gaugino masses, and $S$). Eliminating only thirteen different terms using twenty equations would result in $20-13=7$ RG invariant quantities for this simplified model, on top of $I_1,\ldots,I_6$. 

The ``traced'' $\beta$-functions for this simplified model are again given by eqs. \eqref{TrQ} - \eqref{mhd2} and, as we discussed in the previous subsection, give rise to three independent RG invariants of the form \eqref{I}. Thus, the invariants $I_7,I_8,I_9$ are also invariants in this model. How can we construct the remaining four invariants? Can we again benefit from symmetry arguments?

As we have taken all family space matrices diagonal, there is no longer any flavor mixing present. In other words, the three (s)fermionic generations have completely decoupled. This means that the baryon and lepton numbers are separately conserved for each generation, i.e. $B$ and $L$ can now be split up into $B_1,B_2,B_3$ and $L_1,L_2,L_3$ respectively.\footnote{Note that two linear combinations of these six quantum numbers are equivalent to $B$ and $L$, namely the combinations $B_1+B_2+B_3$ and $L_1+L_2+L_3$ respectively.} {Table \ref{quantum numbers dMSSM}} provides all quantum numbers that pertain to (independent) $\text{U}(1)$ symmetries of the dMSSM. The completeness of this list can be shown in the same way as we did for the MSSM: from table \ref{quantum numbers dMSSM} and appendix \ref{A.2} we infer that we have seventeen gauge multiplets that are subject to three constraints per generation. Hence there must be $17-3\times3=8$ independent $\text{U}(1)$ symmetries in this model. \\
\begin{table}[!th]
\begin{center}
{\renewcommand{\arraystretch}{1.2}
\begin{tabular}{|c|c|c|c|c|c|c|c|c|}
\hline
\textbf{Spin 0} & $\b{Y}$ & $\b{B_1}$ & $\b{B_2}$ & $\b{B_3}$ & $\b{L_1}$ & $\b{L_2}$ & $\b{L_3}$ & $\b{X}$ \\
\hline \hline
$(\wti{Q}_L)_1$ & $\frac{1}{6}$ & $\frac{1}{3}$ & $0$ & $0$ & $0$ & $0$ & $0$ & $1$ \\ 
$(\wti{Q}_L)_2$ & $\frac{1}{6}$ & $0$ & $\frac{1}{3}$ & $0$ & $0$ & $0$ & $0$ & $1$ \\ 
$(\wti{Q}_L)_3$ & $\frac{1}{6}$ & $0$ & $0$ & $\frac{1}{3}$ & $0$ & $0$ & $0$ & $1$ \\ \hdashline
$(\wti{L}_L)_1$ & $-\frac{1}{2}$ & $0$ & $0$ & $0$ & $1$ & $0$ & $0$ & $1$ \\ 
$(\wti{L}_L)_2$ & $-\frac{1}{2}$ & $0$ & $0$ & $0$ & $0$ & $1$ & $0$ & $1$ \\ 
$(\wti{L}_L)_3$ & $-\frac{1}{2}$ & $0$ & $0$ & $0$ & $0$ & $0$ & $1$ & $1$ \\ \hdashline
$(\wti{u}_R^*)_1$ & $-\frac{2}{3}$ & $-\frac{1}{3}$ & $0$ & $0$ & $0$ & $0$ & $0$ & $1$ \\
$(\wti{u}_R^*)_2$ & $-\frac{2}{3}$ & $0$ & $-\frac{1}{3}$ & $0$ & $0$ & $0$ & $0$ & $1$ \\
$(\wti{u}_R^*)_3$ & $-\frac{2}{3}$ & $0$ & $0$ & $-\frac{1}{3}$ & $0$ & $0$ & $0$ & $1$ \\ \hdashline
$(\wti{d}_R^*)_1$ & $\frac{1}{3}$ & $-\frac{1}{3}$ & $0$ & $0$ & $0$ & $0$ & $0$ & $1$ \\
$(\wti{d}_R^*)_2$ & $\frac{1}{3}$ & $0$ & $-\frac{1}{3}$ & $0$ & $0$ & $0$ & $0$ & $1$ \\
$(\wti{d}_R^*)_3$ & $\frac{1}{3}$ & $0$ & $0$ & $-\frac{1}{3}$ & $0$ & $0$ & $0$ & $1$ \\ \hdashline
$(\wti{e}_R^*)_1$ & $1$ & $0$ & $0$ & $0$ & $-1$ & $0$ & $0$ & $1$ \\ 
$(\wti{e}_R^*)_2$ & $1$ & $0$ & $0$ & $0$ & $0$ & $-1$ & $0$ & $1$ \\ 
$(\wti{e}_R^*)_3$ & $1$ & $0$ & $0$ & $0$ & $0$ & $0$ & $-1$ & $1$ \\ \hdashline
$H_u$ & $\frac{1}{2}$ & $0$ & $0$ & $0$ & $0$ & $0$ & $0$ & $-2$ \\
$H_d$ & $-\frac{1}{2}$ & $0$ & $0$ & $0$ & $0$ & $0$ & $0$ & $-2$ \\
\hline
\end{tabular}}
\caption{All quantum numbers that pertain to $\text{U}(1)$ symmetries of the dMSSM family sector. The dashed lines separate the different family multiplets.}
\label{quantum numbers dMSSM}
\end{center}
\end{table}

Since there are four new independent symmetries in the dMSSM with respect to the MSSM, four additional invariants of the form
\begin{equation}
I_Q = \sum_\phi Q_\phi m_\phi^2 - \frac{1}{11} \sum_\phi Y_\phi Q_\phi S + \frac{2}{11} \sum_\phi Y_\phi^2 Q_\phi |M_1|^2 + \frac{3}{2} \sum_d Q_d |M_2|^2 - \frac{8}{9} \sum_t Q_t |M_3|^2
\label{I_Q}
\end{equation}
can be constructed. Clearly, the quantum numbers $B_1-B_2$, $B_1-B_3$, $L_1-L_2$, and $L_1-L_3$ automatically eliminate both $S$ and all gaugino masses. We define
\begin{subequations}
\begin{align}
I_{10} \equiv& \,\sum_\phi ({B_1}_\phi-{B_2}_\phi) m_\phi^2 = 2 m_{\wti{Q}_1}^2 - m_{\wti{u}_1}^2 - m_{\wti{d}_1}^2 - 2 m_{\wti{Q}_2}^2 + m_{\wti{u}_2}^2 + m_{\wti{d}_2}^2 , \\
I_{11} \equiv& \,\sum_\phi ({B_1}_\phi-{B_3}_\phi) m_\phi^2 = 2 m_{\wti{Q}_1}^2 - m_{\wti{u}_1}^2 - m_{\wti{d}_1}^2 - 2 m_{\wti{Q}_3}^2 + m_{\wti{u}_3}^2 + m_{\wti{d}_3}^2 , \\
I_{12} \equiv& \,\sum_\phi ({L_1}_\phi-{L_2}_\phi) m_\phi^2 = 2 m_{\wti{L}_1}^2 - m_{\wti{e}_1}^2 - 2 m_{\wti{L}_2}^2 + m_{\wti{e}_2}^2 , \\
I_{13} \equiv& \,\sum_\phi ({L_1}_\phi-{L_3}_\phi) m_\phi^2 = 2 m_{\wti{L}_1}^2 - m_{\wti{e}_1}^2 - 2 m_{\wti{L}_3}^2 + m_{\wti{e}_3}^2 .
\end{align}
\end{subequations}

\subsection{The pMSSM}
For the pMSSM, with respect to the dMSSM, there are a couple of additional constraints: the first two generations of sfermions are mass degenerate, and the Yukawa and trilinear coupling matrices only have non-zero entries for the third generation sfermions (cf. appendix \ref{C} for a more extensive discussion of the pMSSM). This means that for the pMSSM we have fifteen equations (ten $\beta$-functions for the sfermion masses, two for the Higgs masses, and three for the gaugino masses) to eliminate seven different structures (three structures coming from the (33)-components of the Yukawa and trilinear coupling matrices, three gaugino masses, and $S$), which should yield $15-7=8$ invariant quantities on top of $I_1,\ldots,I_6$. Indeed, for the pMSSM eight RG invariant quantities have been constructed in \cite{Carena:2010gr,Hetzel:2012bk} that involve scalar and gaugino masses only. Can we construct these invariants also based on our approach involving symmetries?

Again, three independent invariants can be constructed that are of the form \eqref{I}, thus $I_7,I_8,I_9$ trivially carry over to the pMSSM. Of the invariants $I_{10},\ldots,I_{13}$, only $I_{11}$ and $I_{13}$ also exist in the pMSSM (the invariants $I_{10}$ and $I_{12}$ vanish due to the mass degeneracy of the first two sfermionic generations).\footnote{Another way to look at this is that in the pMSSM the quantum numbers $B_1,B_2$ and $L_1,L_2$ are each equivalent to each other, which means that there are only \emph{two} new symmetries (and thus two new invariants) with respect to the MSSM.}

Are there additional symmetries in the pMSSM with respect to the flavor diagonal model of the previous subsection? The first thing to note is that the first and second generations of sfermions are completely identical in the pMSSM, which means that we need to consider twelve gauge multiplets only (ten sfermionic multiplets and two Higgses). As the Yukawa and trilinear coupling matrices only have non-zero entries for the third generation sfermions, no first and second generation sfermions feature in any of the family space interactions. This in turn means that any set of quantum numbers for the first two generations automatically corresponds to a symmetry of the pMSSM. The only constraints on allowed sets of quantum numbers (corresponding to symmetries) arise from the third generation sfermions. Similar to the MSSM, we have three constraints coming from the interaction terms in the superpotential and the soft trilinear couplings. This means that there exist $12-3=9$ independent $\text{U}(1)$ symmetries in the pMSSM family sector. {Table~\ref{quantum numbers pMSSM}} provides the quantum numbers that correspond to nine independent $\text{U}(1)$ symmetries of the pMSSM.\footnote{The quantum numbers in this table, of course, are not unique.} \\
\begin{table}[!th]
\begin{center}
{\renewcommand{\arraystretch}{1.2}
\begin{tabular}{|c|c|c|c|c|c|c|c|c|c|}
\hline
\textbf{Spin 0} & $\b{Y_1}$ & $\b{Y}$ & $\b{B_1}$ & $\b{B_3}$ & $\b{L_1}$ & $\b{L_3}$ & $\b{X_{1\ell}}$ & $\b{X_{1q}}$ & $\b{X}$ \\
\hline \hline
$(\wti{Q}_L)_1$ & $\frac{1}{6}$ & $\frac{1}{6}$ & $\frac{1}{3}$ & $0$ & $0$ & $0$ & $0$ & $1$ & $1$ \\ 
$(\wti{Q}_L)_3$ & $0$ & $\frac{1}{6}$ & $0$ & $\frac{1}{3}$ & $0$ & $0$ & $0$ & $0$ & $1$ \\ \hdashline
$(\wti{L}_L)_1$ & $-\frac{1}{2}$ & $-\frac{1}{2}$ & $0$ & $0$ & $1$ & $0$ & $1$ & $0$ & $1$ \\ 
$(\wti{L}_L)_3$ & $0$ & $-\frac{1}{2}$ & $0$ & $0$ & $0$ & $1$ & $0$ & $0$ & $1$ \\ \hdashline
$(\wti{u}_R^*)_1$ & $-\frac{2}{3}$ & $-\frac{2}{3}$ & $-\frac{1}{3}$ & $0$ & $0$ & $0$ & $0$ & $1$ & $1$ \\
$(\wti{u}_R^*)_3$ & $0$ & $-\frac{2}{3}$ & $0$ & $-\frac{1}{3}$ & $0$ & $0$ & $0$ & $0$ & $1$ \\ \hdashline
$(\wti{d}_R^*)_1$ & $\frac{1}{3}$ & $\frac{1}{3}$ & $-\frac{1}{3}$ & $0$ & $0$ & $0$ & $0$ & $1$ & $1$ \\
$(\wti{d}_R^*)_3$ & $0$ & $\frac{1}{3}$ & $0$ & $-\frac{1}{3}$ & $0$ & $0$ & $0$ & $0$ & $1$ \\ \hdashline
$(\wti{e}_R^*)_1$ & $1$ & $1$ & $0$ & $0$ & $-1$ & $0$ & $1$ & $0$ & $1$ \\
$(\wti{e}_R^*)_3$ & $0$ & $1$ & $0$ & $0$ & $0$ & $-1$ & $0$ & $0$ & $1$ \\ \hdashline
$H_u$ & $0$ & $\frac{1}{2}$ & $0$ & $0$ & $0$ & $0$ & $0$ & $0$ & $-2$ \\
$H_d$ & $0$ & $-\frac{1}{2}$ & $0$ & $0$ & $0$ & $0$ & $0$ & $0$ & $-2$ \\
\hline
\end{tabular}}
\caption{All quantum numbers that pertain to $\text{U}(1)$ symmetries of the pMSSM family sector. The dashed lines separate the different family multiplets.}
\label{quantum numbers pMSSM}
\end{center}
\end{table}

Let us now construct the three remaining invariants in the pMSSM, making use of the newly available symmetries and eq. \eqref{I_Q}. A suitable quantum number to cancel both $S$ and all gaugino masses is $10(B_1-L_1)-8Y_1$. For the other two, let us pick $X_{1\ell}$ and $X_{1q}$. We define
\begin{subequations}
\begin{align}
I_{14} \equiv& \,\sum_\phi (10{B_1}_\phi-10{L_1}_\phi-8{Y_1}_\phi) m_\phi^2 \nonumber \\
=& \;12 m_{\wti{Q}_1}^2 - 12 m_{\wti{L}_1}^2 + 6 m_{\wti{u}_1}^2 - 18 m_{\wti{d}_1}^2 + 2 m_{\wti{e}_1}^2 , \\[1mm]
I_{15} \equiv& \,\sum_\phi {X_{1\ell}}_\phi m_\phi^2 + \frac{2}{11} \sum_\phi Y_\phi^2 {X_{1\ell}}_\phi M_1^2 + \frac{3}{2} \sum_d {X_{1\ell}}_d M_2^2 \nonumber \\
=& \;2 m_{\wti{L}_1}^2 + m_{\wti{e}_1}^2 + \frac{3}{11} M_1^2 + 3 M_2^2 , \\[1mm]
I_{16} \equiv& \,\sum_\phi {X_{1q}}_\phi m_\phi^2 + \frac{2}{11} \sum_\phi Y_\phi^2 {X_{1q}}_\phi M_1^2 + \frac{3}{2} \sum_d {X_{1q}}_d M_2^2 - \frac{8}{9} \sum_t {X_{1q}}_t M_3^2 \nonumber \\
=& \;6 m_{\wti{Q}_1}^2 + 3 m_{\wti{u}_1}^2 + 3 m_{\wti{d}_1}^2 + \frac{1}{3} M_1^2 + 9 M_2^2 - \frac{32}{3} M_3^2 .
\end{align}
\end{subequations}

The invariants found for the pMSSM are consistent with the ones derived in \cite{Carena:2010gr,Hetzel:2012bk}.

\section{RG invariants from computer algebraic techniques} \label{4}
Next, we approach the problem of finding RG invariants from a computer algebraic angle. The goal of this approach is to find a method that can in principle be applied to any set of $\beta$-functions, for any theory. By letting a computer do the heavy lifting, we will not have to rely on any properties of the underlying theory as in the case of the previous method, but we will instead be limited by the available computational power. To develop the method, we consider two specific forms of one-loop invariants before extending the method to higher loop orders.

\subsection{Monomial invariants}
Let us first consider the simple class of \emph{monomial invariants}. A monomial invariant $M$ has the following form:
\begin{equation} \label{4.1}
M=\prod_{i=1}^{n}x_{i}^{a_{i}},
\end{equation}
where $x_i$ (with $i=1,\ldots,n$) are running parameters and $\vv{a}\in\mathbb{Z}^{n}.$ The requirement for RG invariance simply reads
\begin{equation} 
\frac{dM}{dt}=M\sum_{i=1}^{n}\frac{a_{i}\beta(x_{i})}{x_{i}}=0 , 
\end{equation}
from which it follows that for all values of the parameters $x_i$, we must have
\begin{equation} \label{4.2}
\sum_{i=1}^{n}\frac{a_{i}\beta(x_{i})}{x_{i}}=0.
\end{equation}

To see how this works in practice, let us consider a simple toy system for two parameters $x$ and $y$ with
\begin{equation} \label{4.3}
\beta(x)=xy+3xy^{2}, \quad \beta(y)=2y^{2}+6y^{3}.
\end{equation}
From requirement \eqref{4.2}, it follows that 
\begin{equation} \label{4.4}
\frac{a_{x}\beta(x)}{x}+\frac{a_{y}\beta(y)}{y}=\left(a_{x}+2a_{y}\right)y+\left(3a_{x}+6a_{y}\right)y^{2}=0, 
\end{equation}
or in matrix form:
\begin{equation}
\left(\begin{array}{cc}
1 & 2\\
3 & 6
\end{array}\right)\left(\begin{array}{c}
a_{x}\\
a_{y}
\end{array}\right)=0.
\end{equation}
Finding a RG invariant has now reduced to solving for the nullspace of a linear system of equations. We find that $\vv{a} = (2,-1)$ spans the nullspace, which leads to the invariant
\begin{equation}
I\equiv\frac{x^{2}}{y}.
\end{equation}

Note that the existence of only a single invariant already implies
the existence of an infinite amount of invariants, since a product
of invariants is also an invariant. However, each of these invariants must solve eq. \eqref{4.2}, and must therefore be included in the nullspace of the linear system of equations. By finding a basis vector for this nullspace, we are effectively including all of these solutions. A higher-dimensional nullspace would imply the existence of multiple independent invariants. Any products of these invariants are also invariant, but their existence is again implied by the linearity of the problem.

\subsection{Polynomial invariants}
Let us now consider \emph{polynomial invariants}. Such an invariant $P$ has the following general form:
\begin{equation} \label{4.6}
P=\sum_{j=1}^{m}C_{j}\prod_{i=1}^{n}x_{i}^{a_{ij}}.
\end{equation}
The powers $a_{ij}$ of the parameters $x_i$ now in fact form a matrix $\b{a}\in\mathbb{Z}^{n\times m}$. We have also introduced $\vv{C}\in\mathbb{Z}^{m}$ as a vector that contains the coefficients for the separate monomial terms. The invariance requirement amounts to
\begin{equation} \label{4.7}
\frac{dP}{dt}=\sum_{j=1}^{m}C_{j}\prod_{i=1}^{n}x_{i}^{a_{ij}}\left(\sum_{k=1}^{n}\frac{a_{kj}}{x_{k}}\beta(x_{k})\right)=0.
\end{equation}
Unlike the case for monomials, we cannot factorize the invariant itself and we are left with a highly nonlinear equation in both the unknowns $\b{a}$ and $\vv{C}$. In addition to the fact that considering products of invariants yields an infinite set of solutions to eq. \eqref{4.7}, now also linear combinations contribute to this issue. Clearly, a method for finding polynomial invariants must be able to deal with both of these sources for ending up with an infinite number of solutions. 

To fix the issue for products of invariants, we introduce the concept
of \emph{dimensionality}. Let us consider the following toy system for two parameters $x$ and $y$, with
\begin{equation} \label{4.8}
\beta(x)=2xy+10y^{3}, \quad \beta(y)=-x-y^{2}.
\end{equation}
We now assign a dimension to the parameters of this system. If we
set $\dim(x)=2$ and $\dim(y)=1$, it follows that $\dim(\beta(x))=3$ and $\dim(\beta(y))=2$. Thus, we find that for all parameters $x_i$, $\dim(\beta(x_i))-\dim(x_i)=c$, where $c$ is a constant. As a consequence, for any monomial $M$,
\begin{equation} \label{4.9}
\dim\left(\frac{d}{dt}M\right) = \dim(M)+c.
\end{equation}
In particular, if we consider two monomials $M$ and $M'$ with
$\dim(M)\neq\dim(M')$, then eq. \eqref{4.9} implies that $\dim(dM/dt)\neq\dim(dM'/dt)$. As a consequence, the monomial terms in $dM/dt$ must be different from those in $dM'/dt$. Hence, if both $M$ and $M'$ are to be included in an invariant, they must be part of two separate groups of monomials that are separately invariant. But then we are considering a linear combination of invariants, which we are trying to avoid. Therefore, whatever the rest of the method is, no results are lost by considering invariants consisting of monomials of the same dimensionality. In addition, this deals with the issue of having an infinite set of solutions to eq. \eqref{4.7} due to considering products of invariants, since the product of two dimensionful invariants has a dimensionality that is different from the two original invariants. Since multiple dimensionalities can be assigned to the parameters of a theory, the dimensionality of a monomial is in general a vector $\vv{d}$. For a system of $\beta$-functions with $r$ distinct dimensionalities (i.e. $\vv{d} \in \mathbb{Z}^{r}$), we define a set of monomials of the same dimensionality as follows:
\begin{equation} \label{4.10}
\mathcal{M}_p (\vv{d}\,) \equiv \left\{ \prod_{i=1}^{n} x_i^{a_i} \:\left|\:  \vv{a} \in \mathbb{Z}^{n}, \: \forall \ell \in \{1,\ldots,r\} : \:\dim_\ell \left( \prod_{i=1}^{n} x_i^{a_i} \right) \right. = d_\ell, \:0 \leq a_i \leq p \right\}.
\end{equation}
The restriction $0 \leq a_i \leq p$ is simply included to ensure that $\mathcal{M}_{p}(\vv{d}\,)$ is a finite set.\footnote{Such a restriction could also be implemented by for example including negative powers.} 

For the toy system of $x$ and $y$, for example, we have
\begin{equation} \label{4.11}
\mathcal{M}_{4}(4)=\left\{ x^{2},xy^{2},y^{4} \right\} .
\end{equation}
A candidate invariant can now be built by using the monomials in this set: 
\begin{equation} \label{4.12}
P_{\vv{d}\,}(\vv{C})=\sum_{j=1}^{s}C_{j}M_{j} ,
\end{equation}
where $M_j \in \mathcal{M}_p(\vv{d}\,)$, and $s$ denotes the size of $\mathcal{M}_{p}(\vv{d}\,)$. Requirement \eqref{4.7} now gives
\begin{align} \label{4.13}
0 &= C_{1}2x\beta(x)+C_{2}y^{2}\beta(x)+C_{2}2xy\beta(y)+C_{3}4y^{3}\beta(y)\nonumber \\[3mm]
&=\left(10C_{2}-4C_{3}\right)y^{5}+\left(20C_{1}-4C_{3}\right)xy^{3}+\left(4C_{1}-2C_{2}\right)x^{2}y.
\end{align}
As this equality must hold for all values of the parameters $x$ and $y$, we again recognize a linear system of equations:
\begin{equation} \label{4.14}
\left(\begin{array}{ccc}
0 & 10 & -4\\
20 & 0 & -4\\
4 & -2 & 0
\end{array}\right)\left(\begin{array}{c}
C_{1}\\
C_{2}\\
C_{3}
\end{array}\right)=0.
\end{equation}
We find that the vector $\vv{C} = (1,2,5)$ spans the nullspace of this matrix, leading to the invariant
\begin{equation} \label{4.15}
I \equiv x^{2}+2xy^{2}+5y^{4}.
\end{equation}

Note that the reduction to a linear system of equations in the coefficients $C_j$ automatically takes care of the problem of having an infinite amount of solutions to eq. \eqref{4.7} due to linear combinations of invariants. Those linear combinations are in fact just linear combinations of the vectors $\vv{C}$. By finding the nullspace for a system of equations like \eqref{4.14}, we handily deal with all problems of eq. \eqref{4.7} while maintaining a simple procedure that can be performed by a computer with ease. This method has been implemented in a C++ application and can be found in \cite{github}. The program is able to find all invariants of the MSSM (derived in section \ref{3}) within seconds. Additionally, two more invariants have been found for the dMSSM: 
\begin{subequations}
\begin{align}
I_{17} \equiv& \;g'^{\,27501} g^{-31965} g_\text{s}^{25920} \left( y_{u_1} y_{u_2} y_{u_3} \right)^{-3859} \left( y_{d_1} y_{d_2} y_{d_3} \right)^{-21481} \left( y_{e_1} y_{e_2} y_{e_3} \right)^{21538} \mu^{751} \\[3mm]
I_{18} \equiv& \;309 M_1 + 4059 M_2 - 6336 M_3 - 693 \,\text{Tr} \left( \b{a_u} \b{y_u}^{-1} \right) - 495 \,\text{Tr} \left( \b{a_d} \b{y_d}^{-1} \right) \nonumber \\[1mm]
& - 242 \,\text{Tr} \left( \b{a_e} \b{y_e}^{-1} \right) + \frac{2937 b}{\mu}
\end{align}
\end{subequations}
For more details on the implementation of the above method, as well as a further extension, see \cite{Rob}. 

\subsection{Higher loop orders}
The computer algebraic method can easily be extended to higher loop orders. To this end, let us consider the general form of a $\beta$-function for $x_i$ in terms of its different loop contributions:
\begin{equation} \label{4.16}
\beta(x_{i})=\beta^{(1)}(x_i)+\frac{1}{16\pi^{2}}\,\beta^{(2)}(x_{i})+\,\ldots
\end{equation}
Two-loop invariants can therefore be found by considering candidate
invariants of the form
\begin{equation} \label{4.17}
I=I_{1}+\frac{1}{16\pi^{2}}\,I_{2}.
\end{equation}
The derivative of $I$ with respect to $t$ reads
\begin{equation} \label{4.18}
\frac{dI}{dt}=I_{1}^{(1)}+\frac{1}{16\pi^{2}}\left(I_{1}^{(2)}+I_{2}^{(1)}\right)+\frac{1}{\left(16\pi^{2}\right)^{2}}\,I_{2}^{(2)},
\end{equation}
where $I_{1,2}^{(j)}$ is the contribution to the derivative that involves the $j$-th loop order $\beta$-function $\beta^{(j)}(x_{i})$. Moreover, the terms have been grouped by equal powers of the factor $1/(16\pi^2)$. The requirement for RG invariance now reads:
\begin{equation} \label{4.19}
I_{1}^{(1)}=0, \quad I_{1}^{(2)}+I_{2}^{(1)}=0,
\end{equation}
which can be reduced to a linear system of equations, equivalent to the method for one-loop invariants. 

The computer algebraic method has been applied to the one and two-loop $\beta$-functions of the MSSM, the dMSSM, and the pMSSM. All one-loop invariants found are consistent with the ones constructed in section \ref{3}. While the one-loop $\beta$-functions are quite well-known, the two-loop ones are not. They were taken from \cite{betaWebsite} and thoroughly checked against the results of \cite{Martin:1993zk, Yamada:1994id, Jack:1994kd}. For the MSSM, we have found the following two-loop invariant:
\begin{equation}
J_1 \equiv \frac{11 M_2}{g^2} - \frac{1}{16\pi^2} \left( M_1 + 209 M_2 - 88 M_3 + \frac{22 b}{\mu} \right) .
\end{equation}
Note that, apparently, only the one-loop invariant $M_2/g^2$ has a two-loop continuation in the MSSM. For the dMSSM, we have found two additional two-loop invariants:
\begin{subequations}
\begin{align}
J_2 \equiv& \;\frac{363 M_1}{g'^2} + \frac{1}{16\pi^2} \left[ \vphantom{\frac{5907 b}{\mu}} 894 M_1 + 6732 M_2 - 16104 M_3 - 1111 \,\text{Tr} \left( \b{a_u} \b{y_u}^{-1} \right) \right. \nonumber \\ 
& \left. - \,1243 \,\text{Tr} \left( \b{a_d} \b{y_d}^{-1} \right) + \frac{5907 b}{\mu} \right] , \\
J_3 \equiv& \;\frac{11 M_3}{g_\text{s}^2} - \frac{1}{16\pi^2} \left[ 66 M_3 + 11 \,\text{Tr} \left( \b{a_u} \b{y_u}^{-1} \right) + 11 \,\text{Tr} \left( \b{a_d} \b{y_d}^{-1} \right) - \frac{33 b}{\mu} \right] ,
\end{align}
\end{subequations}
and for the pMSSM we have found
\begin{subequations}
\begin{align}
J_4 \equiv& \;\frac{2079 M_1}{g'^2} - \frac{1}{16\pi^2} \left( \vphantom{\frac{1518 b}{\mu}} 2869 M_1 + 1485 M_2 - 13640 M_3 + 3762 A_t + 3498 A_\tau \right. \nonumber \\ 
& \left. - \,\frac{1518 b}{\mu} \right) , \\
J_5 \equiv& \;\frac{693 M_3}{g_\text{s}^2} - \frac{1}{16\pi^2} \left( 227 M_1 + 3861 M_2 - 3586 M_3 - 198 A_t - 330 A_\tau + \frac{1320 b}{\mu} \right) .
\end{align}
\end{subequations}
Due to simplifications in the trilinear sector with respect to the MSSM, the dMSSM and pMSSM have two-loop continuations of the one-loop invariants $M_1/g'^2$ and $M_3/g_\text{s}^2$ as well.

\section{Conclusion and outlook}
We have developed two novel, efficient methods for finding RG invariants. The more theoretically inclined approach links the existence of one-loop invariants to symmetries of the underlying theory. For any given supersymmetric theory that has the same structure as the MSSM, the number of RG invariants that involve scalar masses is equal to the number of $\text{U}(1)$ symmetries of its family sector. The computer algebraic method is able to find invariants at higher loop orders and is applicable to any set of RG equations. Both methods have been applied to the $\beta$-functions of for example the unconstrained MSSM and the pMSSM. For the MSSM, three new invariants at one-loop order, and one new invariant at two-loop order have been found. For the pMSSM we have found three new invariants at two-loop order.

A next step in the development of using RG invariants as probes of high-scale physics, could be the construction of new sum rules for various supersymmetry breaking models. This has been done in \cite{Hetzel:2012bk} for the previously known invariants in the pMSSM, and could now be extended to the (d)MSSM. In case supersymmetry is found, it is unlikely that the entire spectrum will quickly be measured. Even our present knowledge of the Higgs mass and other supersymmetry-sensitive data only serves to constrain a number of parameters, depending on the supersymmetry breaking scenario. However, one of the advantages of using RG invariants to probe high-scale physics is the fact that not all parameters of the theory need to be included, simply because the invariants typically contain only a subset of all the parameters,
and some parameters do not enter at all.

Furthermore, since most invariants are directly linked to symmetries of the underlying theory, there is a certain amount of freedom to choose what parameters to include in the invariants. Thus, one might be able to exclude certain parameters from part of the analysis (i.e. particular sum rules of invariants) by cleverly picking linear combinations of quantum numbers such that those parameters do not appear in the relevant invariants. Of course, it could turn out that physics beyond the Standard Model matches a different (perhaps non-supersymmetric) effective field theory rather than one that is discussed in this paper. Since the RG invariants method is completely general, it could still be used to probe high-scale physics models once the $\beta$-functions of the appropriate effective field theory are known.

\appendix
\section{The MSSM} \label{A}
After we consider a general supersymmetric Lagrangian, we will provide the field content and Lagrangian for the MSSM, its $\text{U}(1)$ symmetries, and some one-loop $\beta$-functions. 

\subsection{A general supersymmetric theory}
In supersymmetric theories the fields are grouped into supermultiplets. A chiral supermultiplet (labeled by $i,j$) consists of a complex scalar field $\phi$ and a left-handed Weyl spinor $\psi$. Each gauge group  (labeled by $v$) with corresponding infinitesimal generators $T^a_{v}$, structure constant $f^{abc}_{v}$, and gauge coupling $g_{v}$, gives rise to a gauge supermultiplet. A gauge supermultiplet consists of real gauge boson fields $A_{v\mu}^a$ and Weyl spinors $\lambda^a_{v}$. The gauge invariant superpotential is denoted by $W$, and we define
\begin{equation}
W^i \equiv \frac{\delta W}{\delta \phi_i} , \quad W^{ij} \equiv \frac{\delta^2 W}{\delta \phi_i \delta \phi_j} .
\end{equation}
Using this notation, a general supersymmetric Lagrangian then has the following form \cite{Martin:1997ns}:
\begin{align}
\mathscr{L}_\text{SUSY} =& - \frac{1}{4} F_{v\mu\nu}^{a} F_v^{a\mu\nu} + i {\lambda_v^a}^\dag \overline{\sigma}^\mu \sca{D_\mu \lambda_v}^a + \sca{D^\mu \phi_i}^\dag \sca{D_\mu \phi_i} + i \psi_i^\dag \overline{\sigma}^\mu D_\mu \psi_i \nonumber \\
& - \sqrt{2} \sca{g_v \phi_i^\dag T_v^a \psi_i \cdot \lambda_v^a + \text{h.c.}} - \left| W^i \right|^2 - \frac{1}{2} \sca{W^{ij} \psi_i \cdot \psi_j + \text{h.c.}} \nonumber \\[1mm]
& - \frac{1}{2} g_v^2 \sca{\phi_i^\dag T_v^a \phi_i}^2 ,
\label{SUSY}
\end{align}
where $\overline{\sigma}^\mu \equiv \left( I,-\sigma^1,-\sigma^2,-\sigma^3 \right)$ involves the unit matrix $I$ and the Pauli matrices $\sigma^{a'}$. The indices of Weyl spinors are raised and lowered by the antisymmetric symbol $\epsilon_{\alpha\beta}$ with non-zero components $\epsilon^{12} = - \epsilon^{21} = - \epsilon_{12} = \epsilon_{21} = 1$. By convention, spinor indices are always contracted diagonally downwards for left-handed Weyl spinors. The dot in eq. \eqref{SUSY} is used to denote the contraction of two Weyl spinors that yields a spin-$0$ singlet.\footnote{To avoid clutter, though, this dot is implicit in the remainder of this appendix.} The field strength tensors are defined by 
\begin{equation}
F_{v\mu\nu}^a \equiv \partial_\mu A_{v\nu}^a - \partial_\nu A_{v\mu}^a - g_v f_v^{abc} A_{v\mu}^b A_{v\nu}^c ,
\end{equation}
and the gauge covariant derivatives read
\begin{subequations}
\begin{align}
\sca{D_\mu \lambda_v}^a =& \,\sca{\partial_\mu \delta^{ac} - g_v f_v^{abc} A_{v\mu}^b} \lambda_v^c , \\
D_\mu \phi_i =& \,\sca{\partial_\mu + i g_v A_{v\mu}^a T_v^a} \phi_i , \\[2mm]
D_\mu \psi_i =& \,\sca{\partial_\mu + i g_v A_{v\mu}^a T_v^a} \psi_i .
\end{align}
\end{subequations}

In the remainder of this appendix, the gauge couplings belonging to the gauge groups $\text{U}(1)$, $\text{SU}(2)$, and $\text{SU}(3)$ are denoted by $g'$, $g$, and $g_\text{s}$ respectively. The completely antisymmetric structure constants are respectively given by $0$, $\epsilon^{a'b'c'}$, and $f^{abc}$ (the primes are used to distinguish $\text{SU}(2)$ from $\text{SU}(3)$ indices). The generators of $\text{SU}(2)$ and $\text{SU}(3)$ are proportional to the Pauli matrices $\sigma^{a'}$ and the Gell-Mann matrices $\lambda^a$ respectively.

\subsection{The MSSM Lagrangian} \label{A.2}
The field content of the unbroken MSSM, including the corresponding gauge group representations, is given by tables \ref{chiral} and \ref{gauge}. To distinguish the notation for Standard Model fields from their supersymmetric partners, the latter receive a tilde. In these tables the right-handed Weyl spinors have been conjugated to bring them in a left-handed form: for a Weyl spinor $\psi$ we define $\psi_R^c \equiv i \sigma^2 \psi_R^*$. \\
\begin{table}[!th]
\begin{center}
{\renewcommand{\arraystretch}{1.2}
\begin{tabular}{|c|c|c|c|}
\hline
\textbf{Name} & \textbf{Spin 0} & \textbf{Spin $\b{1/2}$} & \textbf{SU(3)$_\text{C}$ $\b{\times}$ SU(2)$_\text{L}$ $\b{\times}$ U(1)$_\text{Y}$} \\
\hline \hline
\multirow{2}{*}{sleptons, leptons} & $\wti{L}_L = (\wti{\nu}_L \,\, \wti{e}_L)$ & $L_L = (\nu_L \,\, e_L)$ & $(\b{1},\b{2},-\frac{1}{2})$ \\ 
& $\wti{e}_R^*$ & $e_R^c$ & $(\b{1},\b{1},1)$ \\
\hline
\multirow{3}{*}{squarks, quarks} & $\wti{Q}_L = (\wti{u}_L \,\, \wti{d}_L)$ & $Q_L = (u_L \,\, d_L)$ & $(\b{3},\b{2},\frac{1}{6})$ \\ 
& $\wti{u}_R^*$ & $u_R^c$ & $(\b{\overline{3}},\b{1},-\frac{2}{3})$ \\
& $\wti{d}_R^*$ & $d_R^c$ & $(\b{\overline{3}},\b{1},\frac{1}{3})$ \\
\hline
\multirow{2}{*}{Higgs, Higgsinos} & $H_u = (H_u^+ \,\, H_u^0)$ & $\wti{H}_u = (\wti{H}_u^+ \,\, \wti{H}_u^0)$ & $(\b{1},\b{2},\frac{1}{2})$ \\
& $H_d = (H_d^0 \,\, H_d^-)$ & $\wti{H}_d = (\wti{H}_d^0 \,\, \wti{H}_d^-)$ & $(\b{1},\b{2},-\frac{1}{2})$ \\
\hline
\end{tabular}}
\caption{Chiral supermultiplets of the unbroken MSSM and the corresponding gauge group representations.}
\label{chiral}
\end{center}
\end{table}
\begin{table}[!th]
\begin{center}
{\renewcommand{\arraystretch}{1.2}
\begin{tabular}{|c|c|c|c|}
\hline
\textbf{Name} & \textbf{Spin $\b{1/2}$} & \textbf{Spin 1} & \textbf{SU(3)$_\text{C}$ $\b{\times}$ SU(2)$_\text{L}$ $\b{\times}$ U(1)$_\text{Y}$} \\
\hline \hline
gluinos, gluons & $\wti{g}^a$ & $G^a$ & $(\b{8},\b{1},0)$ \\
winos, W bosons & $\wti{W}^{a'}$ & $W^{a'}$ & $(\b{1},\b{3},0)$ \\
bino, B boson & $\wti{B}$ & $B$ & $(\b{1},\b{1},0)$ \\
\hline
\end{tabular}}
\caption{Gauge supermultiplets of the unbroken MSSM and the corresponding gauge group representations.}
\label{gauge}
\end{center}
\end{table}

Including the phenomenologically motivated requirement of $R$-parity conservation, the MSSM superpotential is given by \cite{Martin:1997ns}:
\begin{align}
W_\text{MSSM} =& - \wti{e}_R^\dag \b{y_e} (\wti{L}_L)^\alpha (H_d)_\alpha + \wti{u}_R^\dag \b{y_u} (\wti{Q}_L)^\alpha (H_u)_\alpha - \wti{d}_R^{\,\dag} \b{y_d} (\wti{Q}_L)^\alpha (H_d)_\alpha \nonumber \\[2mm]
& + \mu (H_u)^\alpha (H_d)_\alpha .
\label{WMSSM}
\end{align}
Note that in this expression all color and family indices have been suppressed. For convenience, we do show explicitly the weak isospin doublet indices that are raised and lowered by $\epsilon_{\alpha\beta}$ (like the Weyl spinor indices). All parameters of the MSSM are defined in table \ref{par in mssm} at the end of this subsection.

The full Lagrangian of the MSSM (including soft supersymmetry breaking terms), split up in parts, is given by \cite{Martin:1997ns}:\footnote{As gauge fixing terms are irrelevant for our analyses, we simply omit those here.}

\begin{itemize}
\item \textbf{Kinetic terms for the gauge supermultiplets and gauge interactions:}
\begin{align}
\left[ - \frac{1}{4} F_{v\mu\nu}^{a} F_v^{a\mu\nu} + i {\lambda_v^a}^\dag \overline{\sigma}^\mu \sca{D_\mu \lambda_v}^a \right]_\text{MSSM} =& - \frac{1}{4} G_{\mu\nu}^a G^{a\mu\nu} + i \wti{g}^{a\dag} \overline{\sigma}^\mu (D_\mu \wti{g})^a \nonumber \\ 
& - \frac{1}{4} W_{\mu\nu}^{a'} W^{a'\mu\nu} + i \wti{W}^{a'\dag} \overline{\sigma}^\mu (D_\mu \wti{W})^{a'} \nonumber \\
& - \frac{1}{4} B_{\mu\nu} B^{\mu\nu} + i \wti{B}^\dag \overline{\sigma}^\mu \partial_\mu \wti{B} ,
\end{align}
where the gauge covariant derivatives for the gauginos read
\begin{subequations}
\begin{align}
(D_\mu \wti{g})^a &= (\partial_\mu \delta^{ac} - g_\text{s} f^{abc} G_\mu^b) \wti{g}^c , \\
(D_\mu \wti{W})^{a'} &= (\partial_\mu \delta^{a'c'} - g \epsilon^{a'b'c'} W_\mu^{b'}) \wti{W}^{c'} .
\end{align}
\end{subequations}
The field strength tensors are given by
\begin{subequations}
\begin{align}
G_{\mu\nu}^a &= \partial_\mu G_\nu^a - \partial_\nu G_\mu^a - g_\text{s} f^{abc} G_\mu^b G_\nu^c , \\
W_{\mu\nu}^{a'} &= \partial_\mu W_\nu^{a'} - \partial_\nu W_\mu^{a'} - g \epsilon^{a'b'c'} W_\mu^{b'} W_\nu^{c'} , \\
B_{\mu\nu} &= \partial_\mu B_\nu - \partial_\nu B_\mu .
\end{align}
\end{subequations}

\item \textbf{Kinetic terms for the chiral supermultiplets and gauge interactions:}
\begin{align}
\left[ \sca{D^\mu \phi_i}^\dag \sca{D_\mu \phi_i} + i \psi_i^\dag \overline{\sigma}^\mu D_\mu \psi_i \right]_\text{MSSM} =& \,(D^\mu \wti{L}_L)^\dag D_\mu \wti{L}_L + i L_L^\dag \overline{\sigma}^\mu D_\mu L_L \nonumber \\
& + (D^\mu \wti{e}_R^*)^\dag D_\mu \wti{e}_R^* + i e_R^{c\dag} \overline{\sigma}^\mu D_\mu e_R^c \nonumber \\
& + (D^\mu \wti{Q}_L)^\dag D_\mu \wti{Q}_L + i Q_L^\dag \overline{\sigma}^\mu D_\mu Q_L \nonumber \\
& + (D^\mu \wti{u}_R^*)^\dag D_\mu \wti{u}_R^* + i u_R^{c\dag} \overline{\sigma}^\mu D_\mu u_R^c \nonumber \\
& + (D^\mu \wti{d}_R^*)^\dag D_\mu \wti{d}_R^* + i d_R^{c\dag} \overline{\sigma}^\mu D_\mu d_R^c \nonumber \\
& + (D^\mu H_u)^\dag D_\mu H_u + i \wti{H}_u^\dag \overline{\sigma}^\mu D_\mu \wti{H}_u \nonumber \\
& + (D^\mu H_d)^\dag D_\mu H_d + i \wti{H}_d^\dag \overline{\sigma}^\mu D_\mu \wti{H}_d ,
\end{align}
where the gauge covariant derivatives for the leptons, quarks and Higgs doublets are given by
\begin{subequations}
\begin{align}
D_\mu L_L &= \sca{\partial_\mu + \frac{1}{2} i g W_\mu^{a'} \sigma^{a'} - \frac{1}{2} i g' B_\mu} L_L , \\
D_\mu e_R^c &= \sca{\partial_\mu + i g' B_\mu \vphantom{\frac{1}{2}}} e_R^c , \\
D_\mu Q_L &= \sca{\partial_\mu + \frac{1}{2} i g_\text{s} G_\mu^a \lambda^a + \frac{1}{2} i g W_\mu^{a'} \sigma^{a'} + \frac{1}{6} i g' B_\mu} Q_L , \\
D_\mu u_R^c &= \sca{\partial_\mu - \frac{1}{2} i g_\text{s} G_\mu^a {\lambda^a}^* - \frac{2}{3} i g' B_\mu} u_R^c , \\
D_\mu d_R^c &= \sca{\partial_\mu - \frac{1}{2} i g_\text{s} G_\mu^a {\lambda^a}^* + \frac{1}{3} i g' B_\mu} d_R^c , \\ 
D_\mu H_u &= \sca{\partial_\mu + \frac{1}{2} i g W_\mu^{a'} \sigma^{a'} + \frac{1}{2} i g' B_\mu} H_u , \\
D_\mu H_d &= \sca{\partial_\mu + \frac{1}{2} i g W_\mu^{a'} \sigma^{a'} - \frac{1}{2} i g' B_\mu} H_d .
\end{align}
\end{subequations}
As superpartners have equal quantum numbers, the covariant derivatives for the sleptons, squarks, and Higgsinos are exactly the same.

\item \textbf{Chiral supermultiplets coupled to gauginos:}
\begin{align}
\left[ - \sqrt{2} \sca{g_v \phi_i^\dag T_v^a \psi_i \cdot \lambda_v^a + \text{h.c.}} \right]_\text{MSSM} =& - \sqrt{2} \left( \frac{1}{2} g \wti{L}_L^\dag \sigma^{a'} L_L \wti{W}^{a'} - \frac{1}{2} g' \wti{L}_L^\dag L_L \wti{B} \right. \nonumber \\
& \left. + \,g' \wti{e}_R^\text{T} e_R^c \wti{B} + \frac{1}{2} g_\text{s} \wti{Q}_L^\dag \lambda^a Q_L \wti{g}^a \right. \nonumber \\
& \left. + \,\frac{1}{2} g \wti{Q}_L^\dag \sigma^{a'} Q_L \wti{W}^{a'} + \frac{1}{6} g' \wti{Q}_L^\dag Q_L \wti{B} \right. \nonumber \\
& \left. - \,\frac{1}{2} g_\text{s} \wti{u}_R^\text{T} {\lambda^a}^* u_R^c \wti{g}^a - \frac{2}{3} g' \wti{u}_R^\text{T} u_R^c \wti{B} \right. \nonumber \\
& \left. - \,\frac{1}{2} g_\text{s} \wti{d}_R^{\,\text{T}} {\lambda^a}^* d_R^c \wti{g}^a + \frac{1}{3} g' \wti{d}_R^{\,\text{T}} d_R^c \wti{B} \right. \nonumber \\
& \left. + \,\frac{1}{2} g H_u^\dag \sigma^{a'} \wti{H}_u \wti{W}^{a'} + \frac{1}{2} g' H_u^\dag \wti{H}_u \wti{B} \right. \nonumber \\
& \left. + \,\frac{1}{2} g H_d^\dag \sigma^{a'} \wti{H}_d \wti{W}^{a'} - \frac{1}{2} g' H_d^\dag \wti{H}_d \wti{B} + \,\text{h.c.} \right) .
\label{chiral sups to gauginos}
\end{align}

\item \textbf{Scalar interactions coming from the ``$\b{F}$-fields'':}
\begin{align}
\left[ - \left| W^i \right|^2 \right]_\text{MSSM} =& - \wti{L}_L^\dag \b{y_e^\dag} \wti{e}_R \wti{e}_R^\dag \b{y_e} \wti{L}_L - \wti{e}_R^\dag \b{y_e} \b{y_e^\dag} \wti{e}_R H_d^\dag H_d \nonumber \\
& + (H_d^*)^\beta (\wti{L}_L^\dag)_\beta \b{y_e^\dag y_e} (\wti{L}_L)^\alpha (H_d)_\alpha - \wti{Q}_L^\dag \b{y_u^\dag} \wti{u}_R \wti{u}_R^\dag \b{y_u} \wti{Q}_L \nonumber \\
& - \wti{u}_R^\dag \b{y_u y_u^\dag} \wti{u}_R H_u^\dag H_u + (H_u^*)^\beta (\wti{Q}_L^\dag)_\beta \b{y_u^\dag y_u} (\wti{Q}_L)^\alpha (H_u)_\alpha \nonumber \\
& - \wti{Q}_L^\dag \b{y_d^\dag} \wti{d}_R \wti{d}_R^{\,\dag} \b{y_d} \wti{Q}_L - \wti{d}_R^{\,\dag} \b{y_d y_d^\dag} \wti{d}_R H_d^\dag H_d \nonumber \\
& + (H_d^*)^\beta (\wti{Q}_L^\dag)_\beta \b{y_d^\dag y_d} (\wti{Q}_L)^\alpha (H_d)_\alpha - \left( \wti{Q}_L^\dag \b{y_d^\dag} \wti{d}_R \wti{e}_R^\dag \b{y_e} \wti{L}_L \right. \nonumber \\
& \left. - \,\wti{u}_R^\dag \b{y_u y_d^\dag} \wti{d}_R H_d^\dag H_u - \mu^* \wti{e}_R^\dag \b{y_e} H_u^\dag \wti{L}_L - \mu^* \wti{u}_R^\dag \b{y_u} H_d^\dag \wti{Q}_L \right. \nonumber \\
& \left. - \,\mu^* \wti{d}_R^{\,\dag} \b{y_d} H_u^\dag \wti{Q}_L + \text{h.c.} \vphantom{\wti{Q}_L^\dag} \right) - |\mu|^2 H_u^\dag H_u - |\mu|^2 H_d^\dag H_d .
\end{align}

\item \textbf{Yukawa couplings:}
\begin{align}
\left[ - \frac{1}{2} \sca{W^{ij} \psi_i \cdot \psi_j + \text{h.c.}} \right]_\text{MSSM} =& \;e_R^\dag \b{y_e} (L_L)^\alpha (H_d)_\alpha + e_R^\dag \b{y_e} (\wti{L}_L)^\alpha (\wti{H}_d)_\alpha \nonumber \\
& + \wti{e}_R^\dag \b{y_e} (L_L)^\alpha (\wti{H}_d)_\alpha - u_R^\dag \b{y_u} (Q_L)^\alpha (H_u)_\alpha \nonumber \\
& - u_R^\dag \b{y_u} (\wti{Q}_L)^\alpha (\wti{H}_u)_\alpha - \wti{u}_R^\dag \b{y_u} (Q_L)^\alpha (\wti{H}_u)_\alpha \nonumber \\
& + d_R^\dag \b{y_d} (Q_L)^\alpha (H_d)_\alpha + d_R^\dag \b{y_d} (\wti{Q}_L)^\alpha (\wti{H}_d)_\alpha \nonumber \\
& + \wti{d}_R^{\,\dag} \b{y_d} (Q_L)^\alpha (\wti{H}_d)_\alpha - \mu (\wti{H}_u)^\alpha (\wti{H}_d)_\alpha + \text{h.c.}
\end{align}

\item \textbf{Four-scalar interactions coming from the ``$\b{D}$-fields'':}
\begin{align}
\left[ - \frac{1}{2} g_v^2 \sca{\phi_i^\dag T_v^a \phi_i}^2 \right]_\text{MSSM} =& - \frac{1}{8} g_\text{s}^2 \left( \wti{Q}_L^\dag \lambda^a \wti{Q}_L - \wti{u}_R^\dag \lambda^a \wti{u}_R - \wti{d}_R^{\,\dag} \lambda^a \wti{d}_R \right)^2 \nonumber \\
& - \frac{1}{8} g^2 \left( \wti{L}_L^\dag \sigma^{a'} \wti{L}_L + \wti{Q}_L^\dag \sigma^{a'} \wti{Q}_L + H_u^\dag \sigma^{a'} H_u + H_d^\dag \sigma^{a'} H_d \right)^2 \nonumber \\
& - \frac{1}{2} g'^2 \left( \frac{1}{2} \wti{L}_L^\dag \wti{L}_L - \wti{e}_R^\dag \wti{e}_R - \frac{1}{6} \wti{Q}_L^\dag \wti{Q}_L + \frac{2}{3} \wti{u}_R^\dag \wti{u}_R \right. \nonumber \\
& \left. - \,\frac{1}{3} \wti{d}_R^{\,\dag} \wti{d}_R - \frac{1}{2} H_u^\dag H_u + \frac{1}{2} H_d^\dag H_d \right)^2 .
\end{align}

\item \textbf{Soft supersymmetry breaking terms:}
\begin{align}
\mathscr{L}_\text{MSSM}^\text{soft} =& - \frac{1}{2} \sca{M_3 \wti{g}^{a\text{T}} \wti{g}^a + M_2 \wti{W}^{a'\text{T}} \wti{W}^{a'} + M_1 \wti{B}^\text{T} \wti{B} + \text{h.c.}} \nonumber \\
& + \left[ \wti{e}_R^\dag \b{a_e} (\wti{L}_L)^\alpha (H_d)_\alpha - \wti{u}_R^\dag \b{a_u} (\wti{Q}_L)^\alpha (H_u)_\alpha + \wti{d}_R^{\,\dag} \b{a_d} (\wti{Q}_L)^\alpha (H_d)_\alpha + \text{h.c.} \right] \nonumber \\
& - \wti{L}_L^\dag \b{m_{\wti{L}}^2} \wti{L}_L - \wti{e}_R^\dag \b{m_{\wti{e}}^2} \wti{e}_R - \wti{Q}_L^\dag \b{m_{\wti{Q}}^2} \wti{Q}_L - \wti{u}_R^\dag \b{m_{\wti{u}}^2} \wti{u}_R - \wti{d}_R^{\,\dag} \b{m_{\wti{d}}^2} \wti{d}_R \nonumber \\
& - m_{H_u}^2 H_u^\dag H_u - m_{H_d}^2 H_d^\dag H_d - \left[ b (H_u)^\alpha (H_d)_\alpha + \text{h.c.} \right] .
\label{mssm soft}
\end{align}
\end{itemize}

All parameters of the MSSM are listed in table \ref{par in mssm}. Many degrees of freedom are unphysical though, as they can be absorbed by clever field redefinitions.\footnote{The total number of \emph{independent} parameters in the MSSM is $123$ (not including the strong CP violating angle).} \\
\begin{table}[!th]
\begin{center}
{\renewcommand{\arraystretch}{1.2}
\begin{tabular}{|l|l|l|r|}
\hline
\textbf{Name} & \textbf{Physics description} & \textbf{Math description} & \textbf{\#} \\ \hline \hline
$g_\text{s}, g, g' $ & gauge couplings & real numbers & $3$ \\ $\b{y_e}, \b{y_u}, \b{y_d}$ & Yukawa coupling matrices & complex $3 \times 3$ matrices & $54$ \\ \hdashline
$M_1, M_2, M_3$ & gaugino masses & complex numbers & $6$ \\
$\b{a_e}, \b{a_u}, \b{a_d}$ & trilinear coupling matrices & complex $3 \times 3$ matrices & $54$ \\
$\b{m_{\wti{L}}^2}, \b{m_{\wti{e}}^2}, \b{m_{\wti{Q}}^2}, \b{m_{\wti{u}}^2}, \b{m_{\wti{d}}^2}$ & sfermion mass matrices & Hermitian $3 \times 3$ matrices & $45$ \\
$m_{H_u}, m_{H_d}$ & Higgs masses & real numbers & $2$ \\
$\mu,b$ & Higgs mixing parameters & complex numbers & $4$ \\
\hline
\multicolumn{3}{|r|}{\textit{Total:}} & \textit{168} \\
\hline
\end{tabular}}
\caption{All parameters of the MSSM including soft supersymmetry breaking terms. The parameters that are listed above the horizontal dashed line also occur in the Standard Model.}
\label{par in mssm}
\end{center}
\end{table}

\subsection{U(1) symmetries} \label{A.3}
The MSSM Lagrangian including soft supersymmetry breaking terms has three $\text{U}(1)$ symmetries: weak hypercharge ($Y$), baryon number ($B$), and lepton number ($L$). The last two symmetries are present as a consequence of imposing $R$-parity conservation. If we ignore the soft supersymmetry breaking terms and set the supersymmetry preserving Higgs mixing parameter $\mu$ to zero, then, as it turns out, there are two more $\text{U}(1)$ symmetries in the MSSM. The corresponding groups, that we will call $\text{U}(1)_A$ and $\text{U}(1)_B$, are equivalent to the Peccei-Quinn ($P$) and $R$ symmetries (see \cite{PhysRevLett.38.1440,PhysRevD.16.1791}). Let $\phi$ denote any field contained in the MSSM, then the transformations under $\text{U}(1)_A$ and $\text{U}(1)_B$ are respectively defined as
\begin{equation}
\phi \quad \longrightarrow \quad e^{i Q_A \omega_A} \phi , \quad \quad \phi \quad \longrightarrow \quad e^{i Q_B \omega_B} \phi .
\end{equation}
The charges $Q_A, Q_B$ depend on the fields and $\omega_A, \omega_B$ are free parameters. Tables \ref{chiral charges} and \ref{gauge charges} list the values of $Q_A, Q_B$ for the fields that make up the chiral and gauge supermultiplets respectively. \\
\begin{table}[!th]
\begin{center}
{\renewcommand{\arraystretch}{1.2}
\begin{tabular}{|c|c|c||c|c|c|}
\hline
\textbf{Spin 0} & $\b{Q_A}$ & $\b{Q_B}$ & \textbf{Spin $\b{1/2}$} & $\b{Q_A}$ & $\b{Q_B}$ \\
\hline \hline
$\wti{L}_L$ & 0 & 1 & $L_L$ & 1 & 0 \\ 
$\wti{e}_R^*$ & 0 & 1 & $e_R^c$ & 1 & 0 \\
$\wti{Q}_L$ & 0 & 1 & $Q_L$ & 1 & 0 \\ 
$\wti{u}_R^*$ & 0 & 1 & $u_R^c$ & 1 & 0 \\
$\wti{d}_R^*$ & 0 & 1 & $d_R^c$ & 1 & 0 \\
$H_u$ & $-2$ & 0 & $\wti{H}_u$ & $-1$ & $-1$ \\
$H_d$ & $-2$ & 0 & $\wti{H}_d$ & $-1$ & $-1$ \\
\hline
\end{tabular}}
\caption{The charges $Q_A, Q_B$ for the fields that make up the chiral supermultiplets.}
\label{chiral charges}
\end{center}
\end{table}
\begin{table}[!th]
\begin{center}
{\renewcommand{\arraystretch}{1.2}
\begin{tabular}{|c|c|c||c|c|c|}
\hline
\textbf{Spin $\b{1/2}$} & $\b{Q_A}$ & $\b{Q_B}$ & \textbf{Spin 1} & $\b{Q_A}$ & $\b{Q_B}$ \\
\hline \hline
$\wti{g}^a$ & $-1$ & 1 & $G^a$ & 0 & 0 \\ 
$\wti{W}^{a'}$ & $-1$ & 1 & $W^{a'}$ & 0 & 0 \\
$\wti{B}$ & $-1$ & 1 & $B$ & 0 & 0 \\
\hline
\end{tabular}}
\caption{The charges $Q_A, Q_B$ for the fields that make up the gauge supermultiplets.}
\label{gauge charges}
\end{center}
\end{table}

Now let us consider the MSSM in full (including the soft supersymmetry breaking terms), and let us apply the $\text{U}(1)_A$ and $\text{U}(1)_B$ transformations to all fields. If we would require invariance of the MSSM Lagrangian under these transformations, then we would need to simultaneously redefine the Higgs mixing parameters as follows:\footnote{As the transformations under $\text{U}(1)_A$ and $\text{U}(1)_B$ are parametrized by two independent parameters $\omega_A, \omega_B$, two objects in eqs. \eqref{mu and b} and \eqref{M and a} can each get one of their phases removed by fixing $\omega_A, \omega_B$ in a clever way. Conventionally, the soft breaking parameters $b$ and $M_3$ are made real in this way.}
\begin{equation}
\mu \quad \longrightarrow \quad e^{2 i (\omega_A + \omega_B)} \mu , \quad \quad b \quad \longrightarrow \quad e^{4 i \omega_A} b ,
\label{mu and b}
\end{equation}
and the gaugino masses and trilinear coupling matrices as
\begin{equation}
M_k \quad \longrightarrow \quad e^{2 i (\omega_A - \omega_B)} M_k , \quad \quad \b{a_\psi} \quad \longrightarrow \quad e^{2 i (\omega_A - \omega_B)} \b{a_\psi} ,
\label{M and a}
\end{equation}
where $k = 1,2,3$ and $\psi = e,u,d$. From eq. \eqref{M and a} we infer that the \emph{family sector of the MSSM} (which does not include $\mu$ and $b$), is not invariant under $\text{U}(1)_A$ and $\text{U}(1)_B$ separately, but only under the combination $\text{U}(1)_{A+B}$.\footnote{One may argue that $\mu$ \emph{is} contained in the family sector of the theory; this is a matter of taste. However, since $\mu$ is a supersymmetry preserving parameter, it turns out to be completely irrelevant for our analyses in section \ref{3}. Hence, we can safely ignore this parameter and simply \emph{define} the MSSM family sector to not include $\mu$.} This means that on top of $Y,B,L$, there is a fourth quantum number $X \equiv Q_A+Q_B$ that corresponds to a $\text{U}(1)$ symmetry of the MSSM family sector. 

\subsection{One-loop $\beta$-functions in the MSSM} \label{A.4}
The one-loop $\beta$-functions for the gauge couplings are given by
\begin{equation}
\beta(g') = 11 g'^3 , \quad \beta(g) = g^3 , \quad \beta(g_\text{s}) = -3 g_\text{s}^3 ,
\end{equation}
and those for the squares of the gaugino masses read \cite{Martin:1993zk}
\begin{equation}
\beta \left( |M_1|^2 \right) = 44 g'^2 |M_1|^2 , \quad  \beta \left( |M_2|^2 \right) = 4 g^2 |M_2|^2 , \quad \beta \left( |M_3|^2 \right) = - 12 g_\text{s}^2 |M_3|^2 .
\label{M's}
\end{equation}
The one-loop $\beta$-functions for the soft scalar masses are given by
\begin{subequations}
\begin{align}
\beta(\b{m_{\wti{Q}}^2}) =& - 8 Y_{\wti{Q}_L}^2 g'^2 |M_1|^2 - 6 g^2 |M_2|^2 - \frac{32}{3} g_\text{s}^2 |M_3|^2 + 2 m_{H_u}^2 \b{y_u^\dag y_u} + 2 m_{H_d}^2 \b{y_d^\dag y_d} \nonumber \\
& + \b{m_{\wti{Q}}^2 y_u^\dag y_u} + \b{m_{\wti{Q}}^2 y_d^\dag y_d} + \b{y_u^\dag y_u m_{\wti{Q}}^2} + \b{y_d^\dag y_d m_{\wti{Q}}^2} + 2 \b{y_u^\dag m_{\wti{u}}^2 y_u} \nonumber \\
& + 2 \b{y_d^\dag m_{\wti{d}}^2 y_d} + 2 \b{a_u^\dag a_u} + 2 \b{a_d^\dag a_d} + 2 Y_{\wti{Q}_L} g'^2 S , \\[1mm]
\beta(\b{m_{\wti{L}}^2}) =& - 8 Y_{\wti{L}_L}^2 g'^2 |M_1|^2 - 6 g^2 |M_2|^2 + 2 m_{H_d}^2 \b{y_e^\dag y_e} + \b{m_{\wti{L}}^2 y_e^\dag y_e} + \b{y_e^\dag y_e m_{\wti{L}}^2} \nonumber \\
& + 2 \b{y_e^\dag m_{\wti{e}}^2 y_e} + 2 \b{a_e^\dag a_e} + 2 Y_{\wti{L}_L} g'^2 S , \\[1mm]
\beta(\b{m_{\wti{u}}^2}) =& - 8 Y_{\wti{u}_R^*}^2 g'^2 |M_1|^2 - \frac{32}{3} g_\text{s}^2 |M_3|^2 + 4 m_{H_u}^2 \b{y_u y_u^\dag} + 2 \b{m_{\wti{u}}^2 y_u y_u^\dag} + 2 \b{y_u y_u^\dag m_{\wti{u}}^2} \nonumber \\
& + 4 \b{y_u m_{\wti{Q}}^2 y_u^\dag} + 4 \b{a_u a_u^\dag} + 2 Y_{\wti{u}_R^*} g'^2 S , \\[1mm]
\beta(\b{m_{\wti{d}}^2}) =& - 8 Y_{\wti{d}_R^*}^2 g'^2 |M_1|^2 - \frac{32}{3} g_\text{s}^2 |M_3|^2 + 4 m_{H_d}^2 \b{y_d y_d^\dag} + 2 \b{m_{\wti{d}}^2 y_d y_d^\dag} + 2 \b{y_d y_d^\dag m_{\wti{d}}^2} \nonumber \\
& + 4 \b{y_d m_{\wti{Q}}^2 y_d^\dag} + 4 \b{a_d a_d^\dag} + 2 Y_{\wti{d}_R^*} g'^2 S , \\[1mm]
\beta(\b{m_{\wti{e}}^2}) =& - 8 Y_{\wti{e}_R^*}^2 g'^2 |M_1|^2 + 4 m_{H_d}^2 \b{y_e y_e^\dag} + 2 \b{m_{\wti{e}}^2 y_e y_e^\dag} + 2 \b{y_e y_e^\dag m_{\wti{e}}^2} + 4 \b{y_e m_{\wti{L}}^2 y_e^\dag} \nonumber \\
& + 4 \b{a_e a_e^\dag} + 2 Y_{\wti{e}_R^*} g'^2 S , \\[1mm]
\beta(m_{H_u}^2) =& - 8 Y_{H_u}^2 g'^2 |M_1|^2 - 6 g^2 |M_2|^2 + 6 \,\text{Tr} \Big( m_{H_u}^2 \b{y_u^\dag y_u} + \b{m_{\wti{Q}}^2 y_u^\dag y_u} + \b{y_u^\dag m_{\wti{u}}^2 y_u} \nonumber \\
& + \b{a_u^\dag a_u} \Big) + 2 Y_{H_u} g'^2 S , \label{mhu} \\[1mm]
\beta(m_{H_d}^2) =& - 8 Y_{H_d}^2 g'^2 |M_1|^2 - 6 g^2 |M_2|^2 + 2 \,\text{Tr} \Big( 3 m_{H_d}^2 \b{y_d^\dag y_d} + 3 \b{m_{\wti{Q}}^2 y_d^\dag y_d} + m_{H_d}^2 \b{y_e^\dag y_e} \nonumber \\
& + \b{m_{\wti{L}}^2 y_e^\dag y_e} + 3 \b{y_d^\dag m_{\wti{d}}^2 y_d} + \b{y_e^\dag m_{\wti{e}}^2 y_e} + 3 \b{a_d^\dag a_d} + \b{a_e^\dag a_e} \Big) + 2 Y_{H_d} g'^2 S . \label{mhd}
\end{align}
\end{subequations}

The quantity $S$ that appears in the $\beta$-functions above arises from tadpole diagrams and is defined as
\begin{equation}
S \equiv \;\text{Tr} \bigg( \sum_\phi Y_\phi \b{m_\phi^2} \bigg) = \;\text{Tr} \left( \b{m_{\wti{Q}}^2} - \b{m_{\wti{L}}^2} - 2 \b{m_{\wti{u}}^2} + \b{m_{\wti{d}}^2} + \b{m_{\wti{e}}^2} \right) + m_{H_u}^2 - m_{H_d}^2 .
\label{S}
\end{equation}
The $\beta$-function for $S$ follows directly from the $\beta$-functions for the soft scalar masses and is given by
\begin{equation}
\beta(S) = 2 \sum_\phi Y_\phi^2 g'^2 S = 22 g'^2 S .
\label{betaS}
\end{equation}

The $\beta$-functions for the soft scalar masses that are given above are in correspondence with the ones in \cite{Martin:1993zk}. The advantage of our result though, is that these $\beta$-functions are expressed in terms of the weak hypercharges of the fields, which is useful for our symmetry analyses.

\section{The dMSSM} \label{B}
The dMSSM is the ``flavor-diagonal'' version of the MSSM. The simplifications with respect to the MSSM are as follows:
\begin{itemize} 
\item The Hermitian sfermion mass matrices $\b{m_{\wti{f}}^2}$ with $f=L,e,Q,u,d$ are taken diagonal, i.e.
\begin{equation}
\b{m_{\wti{f}}^2} = \left( \begin{array}{ccc}
m_{\wti{f}_1}^2 & 0 & 0 \\
0 & m_{\wti{f}_2}^2 & 0 \\
0 & 0 & m_{\wti{f}_3}^2
\end{array} \right) .
\end{equation}
\item The Yukawa and trilinear coupling matrices $\b{y_\psi}$ and $\b{a_\psi}$ with $\psi=e,u,d$ are taken real and diagonal, i.e.
\begin{equation}
\b{y_\psi} = \left( \begin{array}{ccc}
y_{\psi_1} & 0 & 0 \\
0 & y_{\psi_2} & 0 \\
0 & 0 & y_{\psi_3}
\end{array} \right) , \quad \b{a_\psi} = \left( \begin{array}{ccc}
a_{\psi_1} & 0 & 0 \\
0 & a_{\psi_2} & 0 \\
0 & 0 & a_{\psi_3}
\end{array} \right) .
\end{equation}
\end{itemize}
All parameters of the dMSSM are listed in table \ref{dMSSM parameters}. \\
\begin{table}[!th]
\begin{center}
{\renewcommand{\arraystretch}{1.2}
\begin{tabular}{|l|l|l|r|}
\hline
\textbf{Name} & \textbf{Physics description} & \textbf{Math description} & \textbf{\#} \\ \hline \hline
$g_\text{s}, g, g' $ & gauge couplings & real numbers & $3$ \\ $\b{y_e}, \b{y_u}, \b{y_d}$ & Yukawa coupling matrices & real, diagonal $3 \times 3$ matrices & $9$ \\ \hdashline
$M_1, M_2, M_3$ & gaugino masses & complex numbers & $6$ \\
$\b{a_e}, \b{a_u}, \b{a_d}$ & trilinear coupling matrices & real, diagonal $3 \times 3$ matrices & $9$ \\
$\b{m_{\wti{L}}^2}, \b{m_{\wti{e}}^2}, \b{m_{\wti{Q}}^2}, \b{m_{\wti{u}}^2}, \b{m_{\wti{d}}^2}$ & sfermion mass matrices & real, diagonal $3 \times 3$ matrices & $15$ \\
$m_{H_u}, m_{H_d}$ & Higgs masses & real numbers & $2$ \\
$\mu,b$ & Higgs mixing parameters & complex numbers & $4$ \\
\hline
\multicolumn{3}{|r|}{\textit{Total:}} & \textit{48} \\
\hline
\end{tabular}}
\caption{All parameters of the dMSSM including soft supersymmetry breaking terms. The parameters that are listed above the horizontal dashed line also occur in the Standard Model.}
\label{dMSSM parameters}
\end{center}
\end{table}

\section{The pMSSM} \label{C}
The pMSSM is a heavily simplified version of the MSSM. Some parameters in the MSSM give rise to processes that seem improbable from a phenomenological point of view, such as flavor-changing neutral currents and CP violation beyond experimental bounds. To suppress these possibilities, one usually imposes several constraints. For the pMSSM they are as follows:
\begin{itemize} 
\item The Hermitian sfermion mass matrices $\b{m_{\wti{f}}^2}$ with $f=L,e,Q,u,d$ are taken diagonal, and the first and second generation masses are assumed to be degenerate, i.e. \vspace*{1mm}
\begin{equation}
\b{m_{\wti{f}}^2} = \left( \begin{array}{ccc}
m_{\wti{f}_1}^2 & 0 & 0 \\
0 & m_{\wti{f}_1}^2 & 0 \\
0 & 0 & m_{\wti{f}_3}^2
\end{array} \right) .
\end{equation}
\item The first and second generation Yukawa couplings are neglected and the third components are taken real:
\begin{equation}
\b{y_e} = \left( \begin{array}{ccc}
0 & 0 & 0 \\
0 & 0 & 0 \\
0 & 0 & y_\tau
\end{array} \right) , \quad \b{y_u} = \left( \begin{array}{ccc}
0 & 0 & 0 \\
0 & 0 & 0 \\
0 & 0 & y_t
\end{array} \right) , \quad \b{y_d} = \left( \begin{array}{ccc}
0 & 0 & 0 \\
0 & 0 & 0 \\
0 & 0 & y_b
\end{array} \right) .
\end{equation}
\item The trilinear coupling matrices are taken proportional to the corresponding Yukawa coupling matrices, which implies
\begin{equation}
a_\tau = A_\tau y_\tau , \quad a_t = A_t y_t , \quad a_b = A_b y_b .
\end{equation}
\item The gaugino masses and Higgs mixing parameters are taken real.
\end{itemize}
All parameters of the pMSSM are listed in table \ref{pMSSM parameters}. \\
\begin{table}[!th]
\begin{center}
{\renewcommand{\arraystretch}{1.2}
\begin{tabular}{|l|l|l|r|}
\hline
\textbf{Name} & \textbf{Physics description} & \textbf{Math description} & \textbf{\#} \\ \hline \hline
$g_\text{s}, g, g'$ & gauge couplings & real numbers & $3$ \\ $y_\tau, y_t, y_b$ & Yukawa couplings & real numbers & $3$ \\ \hdashline
$M_1, M_2, M_3$ & gaugino masses & real numbers & $3$ \\
$A_\tau, A_t, A_b$ & trilinear couplings & real numbers & $3$ \\
$\b{m_{\wti{L}}^2}, \b{m_{\wti{e}}^2}, \b{m_{\wti{Q}}^2}, \b{m_{\wti{u}}^2}, \b{m_{\wti{d}}^2}$ & sfermion mass matrices & real, diagonal $3 \times 3$ matrices & $10$ \\
$m_{H_u}, m_{H_d}$ & Higgs masses & real numbers & $2$ \\
$\mu,b$ & Higgs mixing parameters & real numbers & $2$ \\
\hline
\multicolumn{3}{|r|}{\textit{Total:}} & \textit{26} \\
\hline
\end{tabular}}
\caption{All parameters of the pMSSM including soft supersymmetry breaking terms. The parameters that are listed above the horizontal dashed line also occur in the Standard Model.}
\label{pMSSM parameters}
\end{center}
\end{table}

\section{Summary of results} \label{D}
This appendix lists all RG invariants that have been found for the MSSM, the dMSSM, and the pMSSM. The one-loop invariants have been found by both using the approach involving symmetries and by applying the computer algebraic techniques. The two-loop invariants, however, have only been found by the latter.

\subsection{The MSSM}
Tables \ref{Table2} and \ref{Table3} contain all one and two-loop RG invariants respectively that have been found for the (unconstrained) MSSM. \\
\begin{table}[!th]
\begin{center}
{\renewcommand{\arraystretch}{2.2}
\begin{tabular}{|c|c|c|}
\hline
\textbf{\#} & $\b{Q}$ & \textbf{RG invariant} \\
\hline \hline
$1,2,3$ & & $\frac{M_1}{g'^2}$, $\quad \frac{M_2}{g^2}$, $\quad \frac{M_3}{g_\text{s}^2}$ \\ \hline
$4,5$ & & $\frac{1}{g'^2} - \frac{11}{g^2}$, $\quad \frac{3}{g'^2} + \frac{11}{g_\text{s}^2}$ \\ \hline
$6$ & $Y$ & $\frac{1}{g'^2} \left[ \text{Tr} \left( \b{m_{\wti{Q}}^2} - \b{m_{\wti{L}}^2} - 2 \b{m_{\wti{u}}^2} + \b{m_{\wti{d}}^2} + \b{m_{\wti{e}}^2} \right) + m_{H_u}^2 - m_{H_d}^2 \right]$ \\ \hline
$7$ & $11(B-L)-8Y$ & $\text{Tr} \left( 14 \b{m_{\wti{Q}}^2} - 14 \b{m_{\wti{L}}^2} + 5 \b{m_{\wti{u}}^2} - 19 \b{m_{\wti{d}}^2} + 3 \b{m_{\wti{e}}^2} \right) - 8 m_{H_u}^2 + 8 m_{H_d}^2$ \\ \hline
$8$ & $3B+L$ & $\text{Tr} \left( 6 \b{m_{\wti{Q}}^2} + 2 \b{m_{\wti{L}}^2} - 3 \b{m_{\wti{u}}^2} - 3 \b{m_{\wti{d}}^2} - \b{m_{\wti{e}}^2} \right) - \frac{12}{11} |M_1|^2 + 36 |M_2|^2$ \\ \hline
\multirow{2}{*}{$9$} & \multirow{2}{*}{$X$} & $\text{Tr} \left( 6 \b{m_{\wti{Q}}^2} + 2 \b{m_{\wti{L}}^2} + 3 \b{m_{\wti{u}}^2} + 3 \b{m_{\wti{d}}^2} + \b{m_{\wti{e}}^2} \right) - 4 m_{H_u}^2 - 4 m_{H_d}^2$ \\
& & $+ \,\frac{16}{11} |M_1|^2 + 24 |M_2|^2 - 32 |M_3|^2$ \\
\hline
\end{tabular}}
\caption{One-loop RG invariants in the MSSM. The second column, if applicable, lists the quantum number $Q$ that corresponds to the invariant.}
\label{Table2}
\end{center}
\end{table}

\begin{table}[!th]
\begin{center}
{\renewcommand{\arraystretch}{2.2}
\begin{tabular}{|c|c|}
\hline
\textbf{\#} & \textbf{RG invariant} \\
\hline \hline
$1$ & $\frac{11 M_2}{g^2} - \frac{1}{16\pi^2} \left( M_1 + 209 M_2 - 88 M_3 + \frac{22 b}{\mu} \right)$ \\
\hline
\end{tabular}}
\caption{Two-loop RG invariant in the MSSM.}
\label{Table3}
\end{center}
\end{table}

\subsection{The dMSSM}
Tables \ref{Table7} and \ref{Table8} contain all one and two-loop RG invariants respectively that have been found for the dMSSM. \\
\begin{table}[!th]
\begin{center}
{\renewcommand{\arraystretch}{2.2}
\begin{tabular}{|c|c|c|}
\hline
\textbf{\#} & $\b{Q}$ & \textbf{RG invariant} \\
\hline \hline
$1,2,3$ & & $\frac{M_1}{g'^2}$, $\quad \frac{M_2}{g^2}$, $\quad \frac{M_3}{g_\text{s}^2}$ \\ \hline
$4,5$ & & $\frac{1}{g'^2} - \frac{11}{g^2}$, $\quad \frac{3}{g'^2} + \frac{11}{g_\text{s}^2}$ \\ \hline
$6$ & $Y$ & $\frac{1}{g'^2} \left[ \text{Tr} \left( \b{m_{\wti{Q}}^2} - \b{m_{\wti{L}}^2} - 2 \b{m_{\wti{u}}^2} + \b{m_{\wti{d}}^2} + \b{m_{\wti{e}}^2} \right) + m_{H_u}^2 - m_{H_d}^2 \right]$ \\ \hline
$7$ & $11(B-L)-8Y$ & $\text{Tr} \left( 14 \b{m_{\wti{Q}}^2} - 14 \b{m_{\wti{L}}^2} + 5 \b{m_{\wti{u}}^2} - 19 \b{m_{\wti{d}}^2} + 3 \b{m_{\wti{e}}^2} \right) - 8 m_{H_u}^2 + 8 m_{H_d}^2$ \\ \hline
$8$ & $3B+L$ & $\text{Tr} \left( 6 \b{m_{\wti{Q}}^2} + 2 \b{m_{\wti{L}}^2} - 3 \b{m_{\wti{u}}^2} - 3 \b{m_{\wti{d}}^2} - \b{m_{\wti{e}}^2} \right) - \frac{12}{11} |M_1|^2 + 36 |M_2|^2$ \\ \hline
\multirow{2}{*}{$9$} & \multirow{2}{*}{$X$} & $\text{Tr} \left( 6 \b{m_{\wti{Q}}^2} + 2 \b{m_{\wti{L}}^2} + 3 \b{m_{\wti{u}}^2} + 3 \b{m_{\wti{d}}^2} + \b{m_{\wti{e}}^2} \right) - 4 m_{H_u}^2 - 4 m_{H_d}^2$ \\
& & $+ \,\frac{16}{11} |M_1|^2 + 24 |M_2|^2 - 32 |M_3|^2$ \\ \hline
$10$ & $B_1-B_2$ & $2 m_{\wti{Q}_1}^2 - m_{\wti{u}_1}^2 - m_{\wti{d}_1}^2 - 2 m_{\wti{Q}_2}^2 + m_{\wti{u}_2}^2 + m_{\wti{d}_2}^2$ \\ \hline
$11$ & $B_1-B_3$ & $2 m_{\wti{Q}_1}^2 - m_{\wti{u}_1}^2 - m_{\wti{d}_1}^2 - 2 m_{\wti{Q}_3}^2 + m_{\wti{u}_3}^2 + m_{\wti{d}_3}^2$ \\ \hline
$12$ & $L_1-L_2$ & $2 m_{\wti{L}_1}^2 - m_{\wti{e}_1}^2 - 2 m_{\wti{L}_2}^2 + m_{\wti{e}_2}^2$ \\ \hline
$13$ & $L_1-L_3$ & $2 m_{\wti{L}_1}^2 - m_{\wti{e}_1}^2 - 2 m_{\wti{L}_3}^2 + m_{\wti{e}_3}^2$ \\ \hline
\multirow{2}{*}{$14$} & & $g'^{\,27501} g^{-31965} g_\text{s}^{25920} \left( y_{u_1} y_{u_2} y_{u_3} \right)^{-3859} \left( y_{d_1} y_{d_2} y_{d_3} \right)^{-21481}$ \\
& & $\times \left( y_{e_1} y_{e_2} y_{e_3} \right)^{21538} \mu^{751}$ \\ \hline
\multirow{2}{*}{$15$} & & $309 M_1 + 4059 M_2 - 6336 M_3 - 693 \,\text{Tr} \left( \b{a_u} \b{y_u}^{-1} \right)$ \\ 
& & $- \,495 \,\text{Tr} \left( \b{a_d} \b{y_d}^{-1} \right) - 242 \,\text{Tr} \left( \b{a_e} \b{y_e}^{-1} \right) + \frac{2937 b}{\mu}$ \\
\hline
\end{tabular}}
\caption{One-loop RG invariants in the dMSSM. The second column, if applicable, lists the quantum number $Q$ that corresponds to the invariant.}
\label{Table7}
\end{center}
\end{table}

\begin{table}[!th]
\begin{center}
{\renewcommand{\arraystretch}{2.2}
\begin{tabular}{|c|c|}
\hline
\textbf{\#} & \textbf{RG invariant} \\
\hline \hline
\multirow{2}{*}{$1$} & $\frac{363 M_1}{g'^2} + \frac{1}{16\pi^2} \left[ \vphantom{\frac{5907 b}{\mu}} 894 M_1 + 6732 M_2 - 16104 M_3 - 1111 \,\text{Tr} \left( \b{a_u} \b{y_u}^{-1} \right) \right.$ \\
& $\left. - \,1243 \,\text{Tr} \left( \b{a_d} \b{y_d}^{-1} \right) + \frac{5907 b}{\mu} \right]$ \\ \hline
$2$ & $\frac{11 M_2}{g^2} - \frac{1}{16\pi^2} \left( M_1 + 209 M_2 - 88 M_3 + \frac{22 b}{\mu} \right) $ \\ \hline
$3$ & $\frac{11 M_3}{g_\text{s}^2} - \frac{1}{16\pi^2} \left[ 66 M_3 + 11 \,\text{Tr} \left( \b{a_u} \b{y_u}^{-1} \right) + 11 \,\text{Tr} \left( \b{a_d} \b{y_d}^{-1} \right) - \frac{33 b}{\mu} \right]$ \\
\hline
\end{tabular}}
\caption{Two-loop RG invariants in the dMSSM.}
\label{Table8}
\end{center}
\end{table}

\subsection{The pMSSM}
Tables \ref{Table5} and \ref{Table6} contain all RG invariants that have been found for the pMSSM. These invariants are also listed in \cite{Demir:2004aq,Carena:2010gr,Hetzel:2012bk}, but often as different linear combinations. \\
\begin{table}[!th]
\begin{center}
{\renewcommand{\arraystretch}{2.2}
\begin{tabular}{|c|c|c|}
\hline
\textbf{\#} & $\b{Q}$ & \textbf{RG invariant} \\
\hline \hline
$1,2,3$ & & $\frac{M_1}{g'^2}$, $\quad \frac{M_2}{g^2}$, $\quad \frac{M_3}{g_\text{s}^2}$ \\ \hline
$4,5$ & & $\frac{1}{g'^2} - \frac{11}{g^2}$, $\quad \frac{3}{g'^2} + \frac{11}{g_\text{s}^2}$ \\ \hline
$6$ & $Y$ & $\frac{1}{g'^2} \left[ \text{Tr} \left( \b{m_{\wti{Q}}^2} - \b{m_{\wti{L}}^2} - 2 \b{m_{\wti{u}}^2} + \b{m_{\wti{d}}^2} + \b{m_{\wti{e}}^2} \right) + m_{H_u}^2 - m_{H_d}^2 \right]$ \\ \hline
$7$ & $11(B-L)-8Y$ & $\text{Tr} \left( 14 \b{m_{\wti{Q}}^2} - 14 \b{m_{\wti{L}}^2} + 5 \b{m_{\wti{u}}^2} - 19 \b{m_{\wti{d}}^2} + 3 \b{m_{\wti{e}}^2} \right) - 8 m_{H_u}^2 + 8 m_{H_d}^2$ \\ \hline
$8$ & $3B+L$ & $\text{Tr} \left( 6 \b{m_{\wti{Q}}^2} + 2 \b{m_{\wti{L}}^2} - 3 \b{m_{\wti{u}}^2} - 3 \b{m_{\wti{d}}^2} - \b{m_{\wti{e}}^2} \right) - \frac{12}{11} M_1^2 + 36 M_2^2$ \\ \hline
\multirow{2}{*}{$9$} & \multirow{2}{*}{$X$} & $\text{Tr} \left( 6 \b{m_{\wti{Q}}^2} + 2 \b{m_{\wti{L}}^2} + 3 \b{m_{\wti{u}}^2} + 3 \b{m_{\wti{d}}^2} + \b{m_{\wti{e}}^2} \right) - 4 m_{H_u}^2 - 4 m_{H_d}^2$ \\
& & $+ \,\frac{16}{11} M_1^2 + 24 M_2^2 - 32 M_3^2$ \\ \hline
$10$ & $B_1-B_3$ & $2 m_{\wti{Q}_1}^2 - m_{\wti{u}_1}^2 - m_{\wti{d}_1}^2 - 2 m_{\wti{Q}_3}^2 + m_{\wti{u}_3}^2 + m_{\wti{d}_3}^2$ \\ \hline
$11$ & $L_1-L_3$ & $2 m_{\wti{L}_1}^2 - m_{\wti{e}_1}^2 - 2 m_{\wti{L}_3}^2 + m_{\wti{e}_3}^2$ \\ \hline
$12$ & $10(B_1-L_1)-8Y_1$ & $12 m_{\wti{Q}_1}^2 - 12 m_{\wti{L}_1}^2 + 6 m_{\wti{u}_1}^2 - 18 m_{\wti{d}_1}^2 + 2 m_{\wti{e}_1}^2$ \\ \hline
$13$ & $X_{1\ell}$ & $2 m_{\wti{L}_1}^2 + m_{\wti{e}_1}^2 + \frac{3}{11} M_1^2 + 3 M_2^2$ \\ \hline
$14$ & $X_{1q}$ & $6 m_{\wti{Q}_1}^2 + 3 m_{\wti{u}_1}^2 + 3 m_{\wti{d}_1}^2 + \frac{1}{3} M_1^2 + 9 M_2^2 - \frac{32}{3} M_3^2$ \\ \hline
$15$ & & $g'^{\,73} g^{-297} g_\text{s}^{-2816} y_t^{891} y_b^{693} y_\tau^{330} \mu^{-2013}$ \\ \hline
$16$ & & $73 M_1 - 297 M_2 - 2816 M_3 - 891 A_t - 693 A_b - 330 A_\tau + \frac{2013 b}{\mu}$ \\
\hline
\end{tabular}}
\caption{One-loop RG invariants in the pMSSM. The second column, if applicable, lists the quantum number $Q$ that corresponds to the invariant.}
\label{Table5}
\end{center}
\end{table}

\begin{table}[!th]
\begin{center}
{\renewcommand{\arraystretch}{2.2}
\begin{tabular}{|c|c|}
\hline
\textbf{\#} & \textbf{RG invariant} \\
\hline \hline
$1$ & $\frac{2079 M_1}{g'^2} - \frac{1}{16\pi^2} \left( 2869 M_1 + 1485 M_2 - 13640 M_3 + 3762 A_t + 3498 A_\tau - \frac{1518 b}{\mu} \right) $ \\ \hline
$2$ & $\frac{11 M_2}{g^2} - \frac{1}{16\pi^2} \left( M_1 + 209 M_2 - 88 M_3 + \frac{22 b}{\mu} \right) $ \\ \hline
$3$ & $\frac{693 M_3}{g_\text{s}^2} - \frac{1}{16\pi^2} \left( 227 M_1 + 3861 M_2 - 3586 M_3 - 198 A_t - 330 A_\tau + \frac{1320 b}{\mu} \right) $ \\
\hline
\end{tabular}}
\caption{Two-loop RG invariants in the pMSSM.}
\label{Table6}
\end{center}
\end{table}

\newpage
$ $
\newpage
$ $
\newpage
$ $
\newpage
\acknowledgments
T.v.D. acknowledges financial support from the FP7 EU ``Ideas" program QWORK (contract no. 320389).

\bibliographystyle{JHEP}
\bibliography{literature}

\providecommand{\href}[2]{#2}\begingroup\raggedright\begin{thebibliography}{10}

\bibitem{Demir:2004aq}
D.~A. Demir, {\it {Renormalization group invariants in the MSSM and its
  extensions}},  {\em JHEP} {\bf 0511} (2005) 003,
  [\href{http://xxx.lanl.gov/abs/hep-ph/0408043}{{\tt hep-ph/0408043}}].

\bibitem{Carena:2010gr}
M.~Carena, P.~Draper, N.~R. Shah, and C.~E. Wagner, {\it {Determining the
  Structure of Supersymmetry-Breaking with Renormalization Group Invariants}},
  {\em Phys.Rev.} {\bf D82} (2010) 075005,
  [\href{http://xxx.lanl.gov/abs/1006.4363}{{\tt arXiv:1006.4363}}].

\bibitem{Hetzel:2012bk}
J.~Hetzel and W.~Beenakker, {\it {Renormalisation group invariants and sum
  rules: fast diagnostic tools for probing high-scale physics}},  {\em JHEP}
  {\bf 1210} (2012) 176, [\href{http://xxx.lanl.gov/abs/1204.4336}{{\tt
  arXiv:1204.4336}}].

\bibitem{github}
\url{https://github.com/rbvh/RGIsearch}.

\bibitem{Rob}
R.~Verheyen, {\it {Finding Renormalization Group Equations Using Computer
  Algebraic Methods}},  master's thesis, Radboud University Nijmegen, 2014.

\bibitem{betaWebsite}
\url{http://www.liv.ac.uk/~dij/betas/}.

\bibitem{Martin:1993zk}
S.~P. Martin and M.~T. Vaughn, {\it {Two loop renormalization group equations
  for soft supersymmetry breaking couplings}},  {\em Phys.Rev.} {\bf D50}
  (1994) 2282, [\href{http://xxx.lanl.gov/abs/hep-ph/9311340}{{\tt
  hep-ph/9311340}}].

\bibitem{Yamada:1994id}
Y.~Yamada, {\it {Two loop renormalization group equations for soft SUSY
  breaking scalar interactions: Supergraph method}},  {\em Phys.Rev.} {\bf D50}
  (1994) 3537--3545, [\href{http://xxx.lanl.gov/abs/hep-ph/9401241}{{\tt
  hep-ph/9401241}}].

\bibitem{Jack:1994kd}
I.~Jack and D.~Jones, {\it {Soft supersymmetry breaking and finiteness}},  {\em
  Phys.Lett.} {\bf B333} (1994) 372--379,
  [\href{http://xxx.lanl.gov/abs/hep-ph/9405233}{{\tt hep-ph/9405233}}].

\bibitem{Martin:1997ns}
S.~P. Martin, {\it {A Supersymmetry primer}},
  \href{http://xxx.lanl.gov/abs/hep-ph/9709356}{{\tt hep-ph/9709356}}.

\bibitem{PhysRevLett.38.1440}
R.~D. Peccei and H.~R. Quinn, {\it $\mathrm{CP}$ conservation in the presence
  of instantons},  {\em Phys. Rev. Lett.} {\bf 38} (Jun, 1977) 1440--1443.

\bibitem{PhysRevD.16.1791}
R.~D. Peccei and H.~R. Quinn, {\it Constraints imposed by $\mathrm{CP}$
  conservation in the presence of pseudoparticles},  {\em Phys. Rev. D} {\bf
  16} (Sep, 1977) 1791--1797.

\end{thebibliography}\endgroup

\end{document}